\documentclass[pdflatex,sn-mathphys-num]{sn-jnl}

\usepackage{graphicx}%
\usepackage{multirow}%
\usepackage{amsmath,amssymb,amsfonts}%
\usepackage[title]{appendix}%
\usepackage{xcolor}%
\usepackage{textcomp}%
\usepackage{manyfoot}%
\usepackage{booktabs}%
\usepackage{algorithm}%
\usepackage{algorithmicx}%
\usepackage{algpseudocode}%
\usepackage{listings}%

\theoremstyle{thmstyleone}%
\theoremstyle{thmstyletwo}%

\theoremstyle{thmstylethree}%

\usepackage{array}
\usepackage{adjustbox}
\usepackage[numbers,sort&compress]{natbib}

\usepackage{tablefootnote}
\usepackage{textcomp}
\usepackage{minted}
\usepackage{hyperref}
\usepackage{enumitem}           
\usepackage[T1]{fontenc} 
\usepackage{tikz} 
\usepackage{stfloats}
\usepackage[utf8]{inputenc}
\usepackage{soul}
\usepackage{colortbl}
\definecolor{mygray}{rgb}{0.71,0.71,0.71}
\usepackage{balance}
\usepackage[caption=false]{subfig} 

\usepackage{mdframed}

\newcolumntype{P}[1]{>{\centering\arraybackslash}p{#1}}

\newcommand{\ie}{\textit{i.e.},\ }
\newcommand{\eg}{\textit{e.g.},\ }
\newcommand{\etal}{\textit{et al.} }
\newcommand{\etc}{{\em etc.}}

\usepackage{booktabs}

\definecolor{francBlue}{RGB}{64,76,87}

\usepackage[most]{tcolorbox}
\usepackage[parfill]{parskip}

\tcbset{
  parskip/.style={
    before={\par\pagebreak[0]\parindent=0pt},
    after={\parfillskip=0pt\par},
  },
}
\newtcolorbox{resultbox}[1][]{%
    colback=black!3,
    colframe=black!3,
    notitle,
    sharp corners,
    borderline west={2pt}{0pt}{gray!80!black},
    enhanced,
    breakable,
    boxsep=0pt,
    left=4pt,right=2pt,top=2pt,bottom=2pt,
    }

\definecolor{codebg}{rgb}{0.99,0.99,0.99}
\definecolor{hiliteColor}{rgb}{1,0.92549019607,0.6}
\definecolor{tainted}{rgb}{0,1,1}

\newmintinline{python}{fontsize=\normalsize}

\newminted[PythonSourceCode]{python}{
    labelposition=topline,
    frame=single,
    numbersep=2pt,
    fontsize=\tiny,
    xleftmargin=5pt,
    framerule=0.5pt,
    linenos
}

\newminted[JavaSourceCode]{java}{
    labelposition=topline,
    frame=single,
    numbersep=2pt,
    fontsize=\tiny,
    xleftmargin=5pt,
    framerule=0.5pt,
    linenos
}

\definecolor{magnolia}{rgb}{0.97, 0.96, 1.0}
\definecolor{shadecolor}{rgb}{0.97, 0.96, 1.0}

\newmintinline[codePython]{python}{fontsize=\small}
\newmintinline[codeJava]{java}{fontsize=\small}
\newmintinline[snippetJava]{java}{fontsize=\small,bgcolor=magnolia}
\newmintinline[snippetPython]{python}{fontsize=\small,bgcolor=magnolia}
\newmintinline[snippetPerl]{perl}{fontsize=\small,bgcolor=magnolia}

\newcommand{\category}[1]{\textbf{\textit{#1}}}

\setlist{nosep, topsep=0pt, partopsep=0pt, parsep=0pt, itemsep=0pt}

\usepackage{tikz}

\newcommand{\full}{%
  \tikz\fill[black] (0,0) circle (0.11cm);%
}

\newcommand{\half}{%
  \tikz[baseline=-0.6ex]{
    \draw (0,0) circle (0.11cm);
    \fill (0,0) -- ++(90:0.11cm) arc (90:270:0.11cm) -- cycle;
  }%
}
\newcommand{\none}{%
  \tikz\draw (0,0) circle (0.11cm);%
}

\usepackage{pifont}

\begin{document}

\title[Large Language Models for Software Engineering: A Reproducibility Crisis]{Large Language Models for Software Engineering: A Reproducibility Crisis}

\author[1]{\fnm{Mohammed Latif} \sur{Siddiq}}\email{msiddiq3@nd.edu}

\author[1]{\fnm{Arvin} \sur{Islam-Gomes}}\email{aislamg2@nd.edu}

\author[1]{\fnm{Natalie} \sur{Sekerak}}\email{mnsekerak@nd.edu}

\author*[1]{\fnm{Joanna} \sur{C. S. Santos}}\email{joannacss@nd.edu}

\affil*[1]{\orgdiv{Computer Science and Engineering}, \orgname{University of Notre Dame}, \orgaddress{\street{Holy Cross Drive}, \city{Notre Dame}, \postcode{46556}, \state{IN}, \country{USA}}}

\abstract{
Reproducibility is a cornerstone of scientific progress, yet its state in large language model (LLM)-based software engineering (SE) research remains poorly understood. This paper presents the first large-scale, empirical study of reproducibility practices in LLM-for-SE research. We systematically mined and analyzed \textbf{640 papers} published between 2020 and 2025 across premier software engineering, machine learning, and natural language processing venues, extracting structured metadata from publications, repositories, and documentation. Guided by four research questions, we examine (i) the prevalence of reproducibility smells, (ii) how reproducibility has evolved over time, (iii) whether artifact evaluation badges reliably reflect reproducibility quality, and (iv) how publication venues influence transparency practices. Using a taxonomy of seven smell categories: \category{Code and Execution}, \category{Data}, \category{Documentation}, \category{Environment and Tooling}, \category{Versioning}, \category{Model}, and \category{Access and Legal}, we manually annotated all papers and associated artifacts. Our analysis reveals persistent gaps in artifact availability, environment specification, versioning rigor, and documentation clarity, despite modest improvements in recent years and increased adoption of artifact evaluation processes at top SE venues. Notably, we find that badges often signal artifact presence but do not consistently guarantee execution fidelity or long-term reproducibility. Motivated by these findings, we provide actionable recommendations to mitigate reproducibility smells and introduce a \textbf{Reproducibility Maturity Model (RMM)} to move beyond binary artifact certification toward multi-dimensional, progressive evaluation of reproducibility rigor.}


\maketitle

\section{Introduction}

\textbf{Reproducibility} is a core part of scientific inquiry, enabling independent verification and extension of research findings~\cite{stodden2010scientific,stodden2013toward}. In software engineering (SE) research, reproducibility enables independent verification of results, speeds up cumulative knowledge building, and prompts trust in empirical evidence~\cite{anda2008variability}. However, despite its importance, reproducibility remains a persistent challenge in modern machine learning (ML) and natural language processing (NLP) research~\cite{pineau2021improving,ganesan2023towards}. This challenge is further exacerbated in the context of large language models (LLMs), where rapidly evolving frameworks, model dependencies, and legal or access restrictions can limit experimental replicability~\cite{gudibande2023false,liu2023trustworthy}.

The integration of LLMs into SE workflows, ranging from code generation and bug repair to test synthesis and program comprehension, has grown rapidly since the emergence of foundation models such as \textit{GPT-4}, \textit{Codex}, and \textit{Code Llama}~\cite{openai2023gpt4,chen2021evaluating,CodeLLAMA}. This widespread use has produced hundreds of publications exploring LLM-driven software engineering techniques~\cite{hou2024large}. However, unlike traditional empirical SE research, LLM-based studies often rely on closed-source models, volatile APIs, and non-standardized prompts or hyperparameters, making experimental reproduction particularly difficult~\cite{sallou2023breaking}. As a result, both the SE and ML communities have raised concerns about a looming \textit{``reproducibility crisis''}~\cite{pineau2021improving,ganesan2023towards,Liu_2021}.

Reproducibility has long been a topic of concern in empirical SE research, well before the rise of LLMs. Multiple studies have highlighted gaps between claimed and actual reproducibility of SE experiments, stemming from inadequate artifact sharing, inconsistent environment specifications, lack of standardized benchmarks, and insufficient methodological documentation~\cite{moraila2014measuring,sallou2023breaking}. Unlike controlled laboratory sciences, SE research often involves complex, multi-layered software ecosystems where dependency mismatches, library deprecations, and undocumented preprocessing steps can lead to results that are impossible to replicate, even when the research's source code is available. This has led to community initiatives such as artifact evaluation tracks, reproducibility badges, and standardized packaging practices aimed at improving transparency and reusability of research artifacts~\cite{collberg2016repeatability}. However, as argued by Sallou \etal~\cite{sallou2023breaking}, true reproducibility for work with LLMs requires not just artifact availability but also explicit methodological guidance, and proper information disclosure.

Despite this growing concern, there is still no comprehensive, large-scale study systematically examining how reproducibility is practiced and enforced in LLM-for-SE research. Prior work has either focused on reproducibility in general ML research~\cite{stodden2013toward,pineau2021improving} or on specific subareas of software engineering~\cite{collberg2016repeatability}, leaving a gap at the intersection of these two domains. Moreover, while artifact evaluation initiatives in premier SE venues have made notable progress~\cite{liu2024researchartifactssoftwareengineering,merkel2014docker}, their impact on LLM-based research remains poorly understood.

This manuscript addresses this gap by presenting the \textbf{\textit{first systematic, data-driven investigation of reproducibility practices in LLM-for-SE research}}. We thoroughly analyzed \textbf{640 papers} published between 2020 and 2025 across top SE, ML, and NLP venues. Through a mixed-method approach, we designed and applied a \textit{reproducibility smell taxonomy} encompassing seven key categories: \textit{Code and Execution}, \textit{Data}, \textit{Documentation}, \textit{Environment and Tooling}, \textit{Versioning}, \textit{Model}, and \textit{Access and Legal}. We then manually annotated each paper to assess the prevalence of reproducibility smells and compared patterns across time periods. Additionally, we evaluated the extent to which artifact evaluation badges serve as reliable indicators of reproducibility quality. We additionally conducted a manual analysis of how reproducibility practices evolved across publication venues over the years.

\subsection{Contributions}
The contributions of our work are:
\begin{itemize}[leftmargin=18pt]
    \item We provide the first comprehensive analysis of reproducibility practices in LLM-for-SE research, revealing persistent reproducibility smells, such as \textit{access and legal} smells and \textit{code and execution} smells.
    \item We examine how reproducibility has evolved over time and the co-occurrence of the reproducibility smells.
    \item We investigate whether artifact evaluation badges reliably indicate reproducibility quality.
   \item We analyze how publication venues and artifact evaluation processes influence transparency and reproducibility.
   \item We provide actionable recommendations to mitigate reproducibility smells and introduce a \textbf{\textit{Reproducibility Maturity Model (RMM)}} to move beyond binary artifact certification.
\end{itemize}

Our scripts and data associated with this work are publicly available in our replication package~\cite{siddiq2025llm4se}.

\subsection{Manuscript Organization}
This manuscript is organized as follows. Section~\ref{sec:background} provides background on Large Language Models (LLMs) and their role in Software Engineering research. Section~\ref{sec:methodology} outlines our research questions and the methodology used to address them. Section~\ref{sec:results} presents the results for each research question. Section~\ref{sec:discussion} offers an in-depth discussion of temporal trends in reproducibility smells and examines how model openness contributes to these challenges. Section~\ref{sec:reco} delivers actionable recommendations for authors, reviewers, and publication venues to mitigate reproducibility issues, and introduces the Reproducibility Maturity Model (RMM). Section~\ref{sec:threats} discusses threats to validity. Section~\ref{sec:related} reviews related work and compares our contributions within the broader literature. Finally, Section~\ref{sec:conclusion} concludes the paper.

\textbf{\textit{Note}}: Throughout this manuscript, we discuss widespread reproducibility problems observed across the 640 LLM-for-SE papers analyzed in our study. These issues reflect systemic patterns in the broader LLM-for-SE literature. However, we intentionally do not name or single out any specific papers from our corpus when illustrating these problems. Similar to prior critical reviews~\cite{Baltes2022,Stol2016}, our goal is to merely highlight structural challenges and foster community-wide improvements, not to criticize individual researchers. However, we included the detailed data in our reproducibility package.   
\section{Background}
\label{sec:background}

In this section, we provide a conceptual and technical overview of Large Language Models (LLMs) and their role in Software Engineering (SE) research to understand the challenges and opportunities surrounding their reproducibility.

\subsection{Large Language Models (LLMs)}
\textbf{Large Language Models (LLMs)} are deep neural networks trained on vast text corpora using transformer-based architectures~\cite{attention2017}. These models, such as \textsf{GPT-4}~\cite{openai2023gpt4}, \textsf{LLaMA}~\cite{touvron2023llama}, and \textsf{DeepSeek-V2}~\cite{deepseekai2024deepseekv2strongeconomicalefficient}, demonstrate strong capabilities in understanding and generating human-like language, enabling a wide range of tasks through few-shot or zero-shot prompting~\cite{liu2021pre}. LLMs have rapidly become foundational technologies in natural language processing (NLP), powering advances in text generation, summarization, question answering, and even program synthesis.

Several open-source and commercial LLMs have been further trained and fine-tuned for \textit{code-related} tasks, such as code generation. These LLMs, also known as \textbf{code LLMs}, are large language models specialized for understanding and generating source code, enabling a wide range of programming and software engineering applications. Notably, OpenAI’s \textsf{Codex}~\cite{chen2021evaluating} was trained on GitHub code to power tools like GitHub Copilot, enabling intelligent code completion and generation. Meta’s \textsf{Code Llama} series~\cite{roziere2023code} and Salesforce’s \textsf{CodeGen}~\cite{nijkamp2023codegen2} have followed similar trajectories, offering LLMs optimized for programming languages.

\subsection{LLMs in Software Engineering Research}
The integration of LLMs into Software Engineering (SE) research has given rise to a new field of inquiry (``\textbf{LLM-for-SE}'') where models are applied to automate, support, or improve SE tasks~\cite{hou2024large}. Applications span the entire software lifecycle: requirements engineering (\eg intent extraction \cite{hemmat2025research}), software design (\eg UML generation \cite{conrardy2024image}), implementation (\eg code synthesis \cite{sagodi2024methodology}), testing (\eg test case generation \cite{siddiq24unit} and summarization \cite{sun2024source}), maintenance (\eg bug fixing and documentation \cite{meng2024empirical}), and even project management (\eg issue triage \cite{yu2025triangle}). 

Recent studies have shown promising results using LLMs for tasks such as code completion~\cite{izadi2022codefill,kim2021code,svyatkovskiy2021fast}, code search~\cite{codebert}, code summarization~\cite{gao2022m2ts}, and code generation~\cite{chen2021evaluating}. However, these capabilities often depend heavily on prompt engineering, data quality, and fine-tuning strategies, all of which can vary significantly across implementations, making it challenging to reliably reproduce research results. 

\section{Methodology}
\label{sec:methodology}

\begin{figure}[!htbp]
    \centering
    \includegraphics[width=\linewidth]{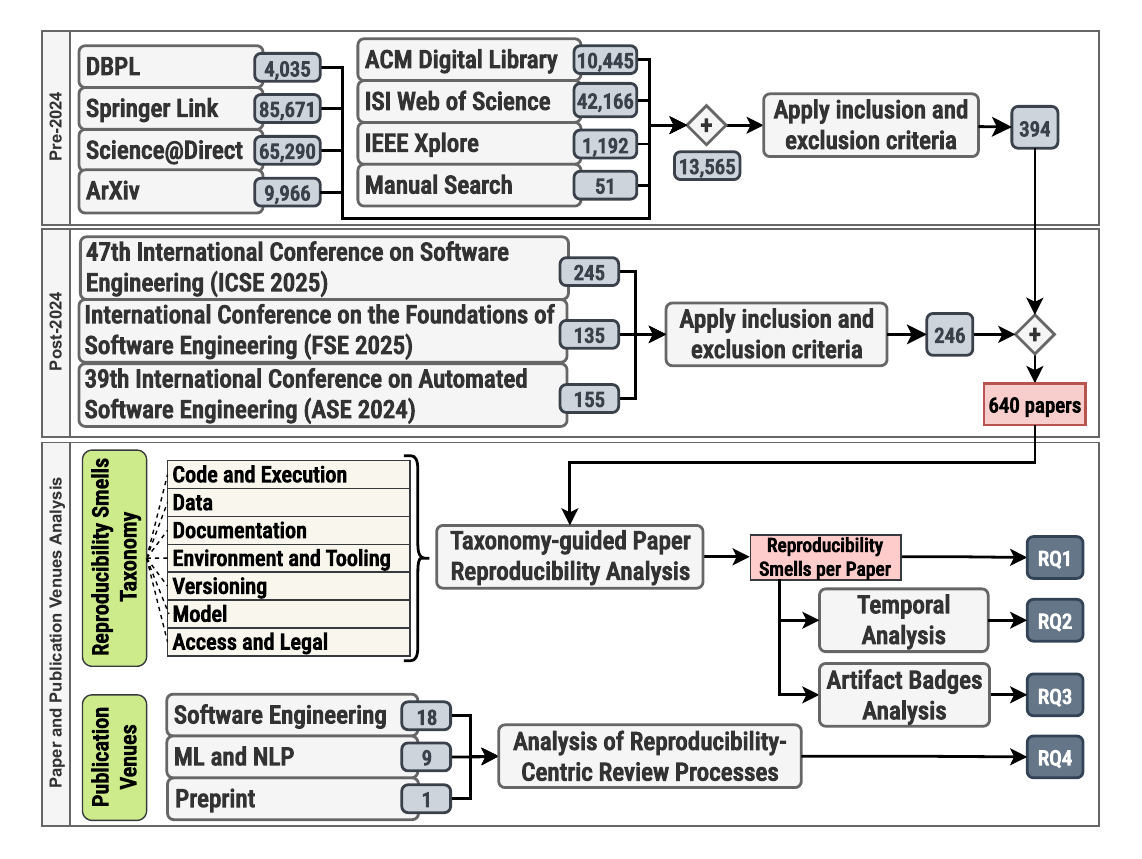}
    \caption{Methodology Overview.}
    \label{fig:method}
\end{figure}

In this section, we outline the methodology we followed to answer our research questions. Figure \ref{fig:method} provides an overview of the methodology of our work.

\subsection{Research Questions}
We answer four research questions in this work:

\begin{itemize}[leftmargin=30pt, label=-,noitemsep,topsep=0pt]
    \item[\textbf{RQ1}] \textit{\textbf{What types of reproducibility smells are most prevalent in LLM-for-SE research?}} 
    \\[3pt]
    This RQ investigates the common patterns of poor reproducibility practices (henceforth, \textbf{\textit{“reproducibility smells”}}) present in published LLM-for-SE studies.
    Identifying prevalent reproducibility smells provides a foundation for understanding systemic weaknesses in current reproducibility practices.
\end{itemize}

\begin{itemize}[leftmargin=30pt, label=-,noitemsep,topsep=0pt]
    \item[\textbf{RQ2}] \textit{\textbf{How has the reproducibility of LLM-for-SE research evolved over time?}} 
    \\[3pt]
    This research question examines temporal trends in reproducibility, focusing on whether more recent studies adopt improved practices compared to earlier work. We examine changes in documentation quality, artifact availability, and adherence to reproducibility checklists to determine whether the field is moving toward more open and verifiable research.
\end{itemize}

\begin{itemize}[leftmargin=30pt, label=-,noitemsep,topsep=0pt]
\item[\textbf{RQ3}] \textit{\textbf{Do artifact evaluation badges reliably reflect the reproducibility of LLM-for-SE research?}}
\\[3pt]
This RQ investigates whether papers awarded artifact evaluation badges exhibit stronger reproducibility characteristics in practice. Specifically, we compare the prevalence of reproducibility smells in papers with badges, examining whether badges serve as a reliable proxy for replication readiness. 
\end{itemize}

\begin{itemize}[leftmargin=30pt, label=-,noitemsep,topsep=0pt]
    \item[\textbf{RQ4}] \textit{\textbf{How do software engineering research venues ensure that published studies are reproducible?}} 
    \\[3pt]
    This RQ investigates the role of conferences and journals in enforcing reproducibility standards. It examines the presence and effectiveness of artifact evaluation tracks, reproducibility badges, mandatory dataset or code submission requirements, review guidelines, and the evolution of review guidelines over time.
\end{itemize}

\subsection{Paper Selection}
To answer our RQs,  we need to collect papers on large language models in software engineering (\ie works on LLM-for-SE). To establish a foundation, we first reviewed the set of papers analyzed in the recent systematic literature review by Hou \etal~\cite{hou2024large}. Since that review covered studies published between January 2017 and January 2024, we further expanded the corpus to include more recent work, ensuring that emerging trends in LLM-for-SE research were represented. The resulting dataset comprises a comprehensive collection of relevant papers spanning both \textbf{pre-2024} (\ie those included in Hou \etal~\cite{hou2024large}) and \textbf{post-2024} publications.  Table~\ref{tab:year_summary} presents the year-wise distribution of the LLM-for-SE papers collected for this study.


\subsubsection{Pre 2024 Papers}
To understand reproducibility practices in published LLM-for-SE, we examined \textbf{394} papers\footnote{Originally, there were 395 papers and one of the papers was retracted.}  from a systematic literature review on the use of LLMs in software engineering research~\cite{hou2024large}. This literature review collected papers published from \textbf{January 2017} to \textbf{January 2024} from a combination of manual searches in top software engineering venues (\eg ICSE, ESEC/FSE, ASE, ISSTA, TOSEM, and TSE) and automated searches across major digital libraries, including IEEE Xplore, ACM Digital Library, ScienceDirect, Web of Science, Springer, arXiv, and DBLP. Moreover, this prior literature review also conducted backward and forward snowballing to identify additional relevant studies through reference lists and citations, ensuring both breadth and depth of coverage. Although their initial search covered papers published between 2017 and 2024, the final set of included papers spanned 2020 to 2024, after applying the inclusion and exclusion criteria summarized in Table \ref{tbl:criteria}.

\begin{table}[!ht]
\caption{Summary of the Inclusion and Exclusion Criteria Used by Hou \etal~\cite{hou2024large}}\label{tbl:criteria}
\footnotesize
\begin{tabular}{@{}ll@{}}
\toprule
\multicolumn{1}{c}{\textbf{Inclusion Criteria}}  & \multicolumn{1}{c}{\textbf{Exclusion Criteria}}  \\ \midrule
\begin{tabular}[c]{@{}p{4cm}@{}}
    \textbf{I1} Written between January 2017 -- January 2024 \\

    \textbf{I2} Focused on a Software Engineering (SE) task\\ 
    \textbf{I3} An LLM is used for an SE task\\ 
    \textbf{I4} Full text is available
\end{tabular} & 
\begin{tabular}[c]{@{}p{8.5cm}@{}}
    \textbf{E1} Short articles (less than 8 pages) \\  
    \textbf{E2} Duplicated articles \\
    \textbf{E3} Position papers, tool demo papers, keynotes, editorials, reviews, tutorials, panel discussions, grey publications, workshop papers, or doctoral symposium papers.\\ 
    \textbf{E4} Studies not in English\\
    \textbf{E5} Article uses LLMs but does not describe the employed techniques.
\end{tabular} \\ \bottomrule
\end{tabular}
\end{table}



\subsubsection{Post 2024 Papers}
For studies published in or after 2024, we focused on the papers published in the research track of the top three (A*) software engineering conferences ranked by the CORE Conference Ranking\footnote{\url{https://portal.core.edu.au/conf-ranks/}}: International Conference on Software Engineering (\textbf{ICSE}), International Conference on Automated Software Engineering (\textbf{ASE}), and International Conference on the Foundations of Software Engineering (\textbf{FSE}). This selection guarantees that all included papers meet the public accessibility requirement (inclusion criterion \textbf{I4}) and do not meet any of the exclusion criteria listed in Table~\ref{tbl:criteria}. To verify the remaining inclusion criteria (\ie~\textbf{I2} and \textbf{I3}), we manually reviewed the full texts of the technical-track papers accepted at \textbf{ICSE 2025}, \textbf{FSE 2025}, and \textbf{ASE 2024}. Through this process, we identified \textbf{102}, \textbf{67}, and \textbf{77} papers, respectively, that employed LLMs for software engineering tasks.

\begin{table}[!htbp]
    \centering
\caption{Year-wise distribution of papers.}
\label{tab:year_summary}

\begin{tabular}{ c         c         c }
    \toprule
\textbf{Year} & \textbf{Number of Papers}& \textbf{List of Papers} \\
\midrule
\textbf{2020} & 7 & \cite{TOSEM-LLM4SE_108,TOSEM-LLM4SE_232,TOSEM-LLM4SE_278,TOSEM-LLM4SE_312,TOSEM-LLM4SE_34,TOSEM-LLM4SE_381,TOSEM-LLM4SE_95} \\
\textbf{2021} & 13 & \cite{TOSEM-LLM4SE_168,TOSEM-LLM4SE_175,TOSEM-LLM4SE_207,TOSEM-LLM4SE_208,TOSEM-LLM4SE_234,TOSEM-LLM4SE_252,TOSEM-LLM4SE_270,TOSEM-LLM4SE_297,TOSEM-LLM4SE_313,TOSEM-LLM4SE_314,TOSEM-LLM4SE_329,TOSEM-LLM4SE_362,TOSEM-LLM4SE_374} \\
\textbf{2022} & 56 & \cite{TOSEM-LLM4SE_101,TOSEM-LLM4SE_109,TOSEM-LLM4SE_125,TOSEM-LLM4SE_140,TOSEM-LLM4SE_143,TOSEM-LLM4SE_144,TOSEM-LLM4SE_151,TOSEM-LLM4SE_152,TOSEM-LLM4SE_186,TOSEM-LLM4SE_201,TOSEM-LLM4SE_205,TOSEM-LLM4SE_206,TOSEM-LLM4SE_211,TOSEM-LLM4SE_216,TOSEM-LLM4SE_249,TOSEM-LLM4SE_250,TOSEM-LLM4SE_251,TOSEM-LLM4SE_253,TOSEM-LLM4SE_257,TOSEM-LLM4SE_259,TOSEM-LLM4SE_260,TOSEM-LLM4SE_269,TOSEM-LLM4SE_277,TOSEM-LLM4SE_281,TOSEM-LLM4SE_291,TOSEM-LLM4SE_295,TOSEM-LLM4SE_296,TOSEM-LLM4SE_3,TOSEM-LLM4SE_31,TOSEM-LLM4SE_317,TOSEM-LLM4SE_32,TOSEM-LLM4SE_324,TOSEM-LLM4SE_33,TOSEM-LLM4SE_335,TOSEM-LLM4SE_339,TOSEM-LLM4SE_341,TOSEM-LLM4SE_343,TOSEM-LLM4SE_35,TOSEM-LLM4SE_354,TOSEM-LLM4SE_357,TOSEM-LLM4SE_364,TOSEM-LLM4SE_366,TOSEM-LLM4SE_373,TOSEM-LLM4SE_38,TOSEM-LLM4SE_382,TOSEM-LLM4SE_386,TOSEM-LLM4SE_39,TOSEM-LLM4SE_41,TOSEM-LLM4SE_53,TOSEM-LLM4SE_59,TOSEM-LLM4SE_65,TOSEM-LLM4SE_80,TOSEM-LLM4SE_93,TOSEM-LLM4SE_94,TOSEM-LLM4SE_96,TOSEM-LLM4SE_98} \\
\textbf{2023} & 272 & \cite{TOSEM-LLM4SE_0,TOSEM-LLM4SE_1,TOSEM-LLM4SE_10,TOSEM-LLM4SE_100,TOSEM-LLM4SE_102,TOSEM-LLM4SE_103,TOSEM-LLM4SE_105,TOSEM-LLM4SE_106,TOSEM-LLM4SE_107,TOSEM-LLM4SE_11,TOSEM-LLM4SE_110,TOSEM-LLM4SE_111,TOSEM-LLM4SE_112,TOSEM-LLM4SE_113,TOSEM-LLM4SE_114,TOSEM-LLM4SE_115,TOSEM-LLM4SE_116,TOSEM-LLM4SE_119,TOSEM-LLM4SE_12,TOSEM-LLM4SE_120,TOSEM-LLM4SE_121,TOSEM-LLM4SE_122,TOSEM-LLM4SE_123,TOSEM-LLM4SE_124,TOSEM-LLM4SE_127,TOSEM-LLM4SE_128,TOSEM-LLM4SE_129,TOSEM-LLM4SE_130,TOSEM-LLM4SE_131,TOSEM-LLM4SE_132,TOSEM-LLM4SE_133,TOSEM-LLM4SE_134,TOSEM-LLM4SE_135,TOSEM-LLM4SE_136,TOSEM-LLM4SE_137,TOSEM-LLM4SE_138,TOSEM-LLM4SE_139,TOSEM-LLM4SE_142,TOSEM-LLM4SE_145,TOSEM-LLM4SE_146,TOSEM-LLM4SE_147,TOSEM-LLM4SE_148,TOSEM-LLM4SE_149,TOSEM-LLM4SE_150,TOSEM-LLM4SE_153,TOSEM-LLM4SE_154,TOSEM-LLM4SE_155,TOSEM-LLM4SE_156,TOSEM-LLM4SE_157,TOSEM-LLM4SE_159,TOSEM-LLM4SE_161,TOSEM-LLM4SE_162,TOSEM-LLM4SE_163,TOSEM-LLM4SE_164,TOSEM-LLM4SE_165,TOSEM-LLM4SE_166,TOSEM-LLM4SE_167,TOSEM-LLM4SE_169,TOSEM-LLM4SE_17,TOSEM-LLM4SE_170,TOSEM-LLM4SE_171,TOSEM-LLM4SE_173,TOSEM-LLM4SE_174,TOSEM-LLM4SE_176,TOSEM-LLM4SE_177,TOSEM-LLM4SE_178,TOSEM-LLM4SE_179,TOSEM-LLM4SE_18,TOSEM-LLM4SE_180,TOSEM-LLM4SE_181,TOSEM-LLM4SE_182,TOSEM-LLM4SE_183,TOSEM-LLM4SE_184,TOSEM-LLM4SE_185,TOSEM-LLM4SE_187,TOSEM-LLM4SE_189,TOSEM-LLM4SE_19,TOSEM-LLM4SE_190,TOSEM-LLM4SE_191,TOSEM-LLM4SE_192,TOSEM-LLM4SE_193,TOSEM-LLM4SE_194,TOSEM-LLM4SE_195,TOSEM-LLM4SE_196,TOSEM-LLM4SE_197,TOSEM-LLM4SE_199,TOSEM-LLM4SE_2,TOSEM-LLM4SE_20,TOSEM-LLM4SE_200,TOSEM-LLM4SE_202,TOSEM-LLM4SE_203,TOSEM-LLM4SE_204,TOSEM-LLM4SE_209,TOSEM-LLM4SE_21,TOSEM-LLM4SE_210,TOSEM-LLM4SE_214,TOSEM-LLM4SE_215,TOSEM-LLM4SE_217,TOSEM-LLM4SE_218,TOSEM-LLM4SE_22,TOSEM-LLM4SE_221,TOSEM-LLM4SE_222,TOSEM-LLM4SE_223,TOSEM-LLM4SE_224,TOSEM-LLM4SE_225,TOSEM-LLM4SE_226,TOSEM-LLM4SE_227,TOSEM-LLM4SE_228,TOSEM-LLM4SE_229,TOSEM-LLM4SE_23,TOSEM-LLM4SE_230,TOSEM-LLM4SE_231,TOSEM-LLM4SE_233,TOSEM-LLM4SE_238,TOSEM-LLM4SE_239,TOSEM-LLM4SE_24,TOSEM-LLM4SE_240,TOSEM-LLM4SE_241,TOSEM-LLM4SE_242,TOSEM-LLM4SE_243,TOSEM-LLM4SE_244,TOSEM-LLM4SE_245,TOSEM-LLM4SE_246,TOSEM-LLM4SE_247,TOSEM-LLM4SE_248,TOSEM-LLM4SE_25,TOSEM-LLM4SE_254,TOSEM-LLM4SE_255,TOSEM-LLM4SE_256,TOSEM-LLM4SE_258,TOSEM-LLM4SE_26,TOSEM-LLM4SE_261,TOSEM-LLM4SE_262,TOSEM-LLM4SE_263,TOSEM-LLM4SE_265,TOSEM-LLM4SE_266,TOSEM-LLM4SE_267,TOSEM-LLM4SE_268,TOSEM-LLM4SE_27,TOSEM-LLM4SE_271,TOSEM-LLM4SE_272,TOSEM-LLM4SE_273,TOSEM-LLM4SE_274,TOSEM-LLM4SE_275,TOSEM-LLM4SE_276,TOSEM-LLM4SE_28,TOSEM-LLM4SE_280,TOSEM-LLM4SE_282,TOSEM-LLM4SE_283,TOSEM-LLM4SE_284,TOSEM-LLM4SE_285,TOSEM-LLM4SE_286,TOSEM-LLM4SE_287,TOSEM-LLM4SE_288,TOSEM-LLM4SE_289,TOSEM-LLM4SE_29,TOSEM-LLM4SE_290,TOSEM-LLM4SE_292,TOSEM-LLM4SE_293,TOSEM-LLM4SE_294,TOSEM-LLM4SE_30,TOSEM-LLM4SE_300,TOSEM-LLM4SE_302,TOSEM-LLM4SE_303,TOSEM-LLM4SE_304,TOSEM-LLM4SE_305,TOSEM-LLM4SE_306,TOSEM-LLM4SE_307,TOSEM-LLM4SE_308,TOSEM-LLM4SE_310,TOSEM-LLM4SE_311,TOSEM-LLM4SE_315,TOSEM-LLM4SE_316,TOSEM-LLM4SE_318,TOSEM-LLM4SE_319,TOSEM-LLM4SE_320,TOSEM-LLM4SE_323,TOSEM-LLM4SE_325,TOSEM-LLM4SE_326,TOSEM-LLM4SE_327,TOSEM-LLM4SE_328,TOSEM-LLM4SE_332,TOSEM-LLM4SE_333,TOSEM-LLM4SE_334,TOSEM-LLM4SE_336,TOSEM-LLM4SE_338,TOSEM-LLM4SE_342,TOSEM-LLM4SE_344,TOSEM-LLM4SE_345,TOSEM-LLM4SE_347,TOSEM-LLM4SE_348,TOSEM-LLM4SE_349,TOSEM-LLM4SE_350,TOSEM-LLM4SE_351,TOSEM-LLM4SE_352,TOSEM-LLM4SE_353,TOSEM-LLM4SE_356,TOSEM-LLM4SE_358,TOSEM-LLM4SE_359,TOSEM-LLM4SE_36,TOSEM-LLM4SE_360,TOSEM-LLM4SE_361,TOSEM-LLM4SE_363,TOSEM-LLM4SE_367,TOSEM-LLM4SE_369,TOSEM-LLM4SE_37,TOSEM-LLM4SE_370,TOSEM-LLM4SE_371,TOSEM-LLM4SE_372,TOSEM-LLM4SE_375,TOSEM-LLM4SE_376,TOSEM-LLM4SE_377,TOSEM-LLM4SE_378,TOSEM-LLM4SE_379,TOSEM-LLM4SE_380,TOSEM-LLM4SE_383,TOSEM-LLM4SE_384,TOSEM-LLM4SE_385,TOSEM-LLM4SE_387,TOSEM-LLM4SE_388,TOSEM-LLM4SE_390,TOSEM-LLM4SE_391,TOSEM-LLM4SE_392,TOSEM-LLM4SE_4,TOSEM-LLM4SE_40,TOSEM-LLM4SE_42,TOSEM-LLM4SE_43,TOSEM-LLM4SE_44,TOSEM-LLM4SE_45,TOSEM-LLM4SE_46,TOSEM-LLM4SE_47,TOSEM-LLM4SE_49,TOSEM-LLM4SE_5,TOSEM-LLM4SE_50,TOSEM-LLM4SE_51,TOSEM-LLM4SE_52,TOSEM-LLM4SE_54,TOSEM-LLM4SE_55,TOSEM-LLM4SE_57,TOSEM-LLM4SE_58,TOSEM-LLM4SE_6,TOSEM-LLM4SE_60,TOSEM-LLM4SE_61,TOSEM-LLM4SE_62,TOSEM-LLM4SE_64,TOSEM-LLM4SE_67,TOSEM-LLM4SE_68,TOSEM-LLM4SE_69,TOSEM-LLM4SE_7,TOSEM-LLM4SE_70,TOSEM-LLM4SE_71,TOSEM-LLM4SE_72,TOSEM-LLM4SE_73,TOSEM-LLM4SE_74,TOSEM-LLM4SE_75,TOSEM-LLM4SE_76,TOSEM-LLM4SE_77,TOSEM-LLM4SE_78,TOSEM-LLM4SE_79,TOSEM-LLM4SE_8,TOSEM-LLM4SE_84,TOSEM-LLM4SE_85,TOSEM-LLM4SE_86,TOSEM-LLM4SE_87,TOSEM-LLM4SE_88,TOSEM-LLM4SE_89,TOSEM-LLM4SE_9,TOSEM-LLM4SE_90,TOSEM-LLM4SE_91,TOSEM-LLM4SE_92,TOSEM-LLM4SE_97,TOSEM-LLM4SE_99} \\
\textbf{2024} & 123 & \cite{ASE-2024_1,ASE-2024_10,ASE-2024_11,ASE-2024_12,ASE-2024_13,ASE-2024_14,ASE-2024_15,ASE-2024_16,ASE-2024_17,ASE-2024_18,ASE-2024_19,ASE-2024_2,ASE-2024_20,ASE-2024_21,ASE-2024_22,ASE-2024_23,ASE-2024_24,ASE-2024_25,ASE-2024_26,ASE-2024_27,ASE-2024_28,ASE-2024_29,ASE-2024_3,ASE-2024_30,ASE-2024_31,ASE-2024_32,ASE-2024_33,ASE-2024_34,ASE-2024_35,ASE-2024_36,ASE-2024_37,ASE-2024_38,ASE-2024_39,ASE-2024_4,ASE-2024_40,ASE-2024_41,ASE-2024_42,ASE-2024_43,ASE-2024_44,ASE-2024_45,ASE-2024_46,ASE-2024_47,ASE-2024_48,ASE-2024_49,ASE-2024_5,ASE-2024_50,ASE-2024_51,ASE-2024_52,ASE-2024_53,ASE-2024_54,ASE-2024_55,ASE-2024_56,ASE-2024_57,ASE-2024_58,ASE-2024_59,ASE-2024_6,ASE-2024_60,ASE-2024_61,ASE-2024_62,ASE-2024_63,ASE-2024_64,ASE-2024_65,ASE-2024_66,ASE-2024_67,ASE-2024_68,ASE-2024_69,ASE-2024_7,ASE-2024_70,ASE-2024_71,ASE-2024_72,ASE-2024_73,ASE-2024_74,ASE-2024_75,ASE-2024_76,ASE-2024_77,ASE-2024_8,ASE-2024_9,TOSEM-LLM4SE_104,TOSEM-LLM4SE_117,TOSEM-LLM4SE_118,TOSEM-LLM4SE_126,TOSEM-LLM4SE_13,TOSEM-LLM4SE_14,TOSEM-LLM4SE_141,TOSEM-LLM4SE_15,TOSEM-LLM4SE_158,TOSEM-LLM4SE_16,TOSEM-LLM4SE_160,TOSEM-LLM4SE_172,TOSEM-LLM4SE_188,TOSEM-LLM4SE_198,TOSEM-LLM4SE_212,TOSEM-LLM4SE_213,TOSEM-LLM4SE_219,TOSEM-LLM4SE_220,TOSEM-LLM4SE_235,TOSEM-LLM4SE_236,TOSEM-LLM4SE_237,TOSEM-LLM4SE_264,TOSEM-LLM4SE_279,TOSEM-LLM4SE_298,TOSEM-LLM4SE_299,TOSEM-LLM4SE_301,TOSEM-LLM4SE_309,TOSEM-LLM4SE_321,TOSEM-LLM4SE_322,TOSEM-LLM4SE_330,TOSEM-LLM4SE_331,TOSEM-LLM4SE_337,TOSEM-LLM4SE_340,TOSEM-LLM4SE_346,TOSEM-LLM4SE_355,TOSEM-LLM4SE_365,TOSEM-LLM4SE_368,TOSEM-LLM4SE_389,TOSEM-LLM4SE_393,TOSEM-LLM4SE_48,TOSEM-LLM4SE_56,TOSEM-LLM4SE_63,TOSEM-LLM4SE_66,TOSEM-LLM4SE_81,TOSEM-LLM4SE_82,TOSEM-LLM4SE_83} \\
\textbf{2025} & 169 & \cite{FSE-2025_1,FSE-2025_10,FSE-2025_11,FSE-2025_12,FSE-2025_13,FSE-2025_14,FSE-2025_15,FSE-2025_16,FSE-2025_17,FSE-2025_18,FSE-2025_19,FSE-2025_2,FSE-2025_20,FSE-2025_21,FSE-2025_22,FSE-2025_23,FSE-2025_24,FSE-2025_25,FSE-2025_26,FSE-2025_27,FSE-2025_28,FSE-2025_29,FSE-2025_3,FSE-2025_30,FSE-2025_31,FSE-2025_32,FSE-2025_33,FSE-2025_34,FSE-2025_35,FSE-2025_36,FSE-2025_37,FSE-2025_38,FSE-2025_39,FSE-2025_4,FSE-2025_40,FSE-2025_41,FSE-2025_42,FSE-2025_43,FSE-2025_44,FSE-2025_45,FSE-2025_46,FSE-2025_47,FSE-2025_48,FSE-2025_49,FSE-2025_5,FSE-2025_50,FSE-2025_51,FSE-2025_52,FSE-2025_53,FSE-2025_54,FSE-2025_55,FSE-2025_56,FSE-2025_57,FSE-2025_58,FSE-2025_59,FSE-2025_6,FSE-2025_60,FSE-2025_61,FSE-2025_62,FSE-2025_63,FSE-2025_64,FSE-2025_65,FSE-2025_66,FSE-2025_67,FSE-2025_7,FSE-2025_8,FSE-2025_9,ICSE-2025_1,ICSE-2025_10,ICSE-2025_100,ICSE-2025_101,ICSE-2025_102,ICSE-2025_103,ICSE-2025_104,ICSE-2025_11,ICSE-2025_12,ICSE-2025_13,ICSE-2025_14,ICSE-2025_15,ICSE-2025_16,ICSE-2025_17,ICSE-2025_18,ICSE-2025_19,ICSE-2025_2,ICSE-2025_20,ICSE-2025_21,ICSE-2025_22,ICSE-2025_23,ICSE-2025_24,ICSE-2025_25,ICSE-2025_26,ICSE-2025_27,ICSE-2025_28,ICSE-2025_29,ICSE-2025_3,ICSE-2025_30,ICSE-2025_31,ICSE-2025_32,ICSE-2025_33,ICSE-2025_34,ICSE-2025_35,ICSE-2025_36,ICSE-2025_37,ICSE-2025_38,ICSE-2025_39,ICSE-2025_4,ICSE-2025_40,ICSE-2025_41,ICSE-2025_42,ICSE-2025_43,ICSE-2025_44,ICSE-2025_45,ICSE-2025_46,ICSE-2025_47,ICSE-2025_48,ICSE-2025_49,ICSE-2025_5,ICSE-2025_50,ICSE-2025_51,ICSE-2025_52,ICSE-2025_53,ICSE-2025_54,ICSE-2025_55,ICSE-2025_56,ICSE-2025_57,ICSE-2025_58,ICSE-2025_59,ICSE-2025_6,ICSE-2025_60,ICSE-2025_61,ICSE-2025_62,ICSE-2025_63,ICSE-2025_64,ICSE-2025_65,ICSE-2025_66,ICSE-2025_67,ICSE-2025_68,ICSE-2025_69,ICSE-2025_7,ICSE-2025_70,ICSE-2025_71,ICSE-2025_72,ICSE-2025_73,ICSE-2025_74,ICSE-2025_75,ICSE-2025_77,ICSE-2025_78,ICSE-2025_79,ICSE-2025_8,ICSE-2025_80,ICSE-2025_81,ICSE-2025_82,ICSE-2025_83,ICSE-2025_84,ICSE-2025_85,ICSE-2025_86,ICSE-2025_87,ICSE-2025_88,ICSE-2025_89,ICSE-2025_9,ICSE-2025_90,ICSE-2025_91,ICSE-2025_92,ICSE-2025_93,ICSE-2025_95,ICSE-2025_96,ICSE-2025_97,ICSE-2025_98,ICSE-2025_99} \\
\bottomrule
\end{tabular}

\end{table}

\subsection{Reproducibility Smell Analysis}
After merging the papers collected pre-2024 and post-2024, we obtained a total of \textbf{640} papers.
To systematically evaluate the reproducibility of LLM-for-SE works, we conducted a structured analysis of each of these 640 selected papers, focusing on the key dimensions that affect experimental replication and verification. As part of this process, we developed \textbf{\textit{a reproducibility smell taxonomy}} to categorize and quantify recurring issues that hinder reproducibility in LLM-for-SE research. Our taxonomy draws conceptual inspiration from Hassan \textit{et.al.}’s systematic literature review of reproducibility debt in scientific software~\cite{HASSAN2025112327}, which identified seven broad categories of issues (data-centric, code-centric, documentation-centric, process-centric, human-centric, legal, and version-centric) and provided a high-level organization of reproducibility-related technical debt across computational sciences.

However, Hassan \textit{et.al.}’s framework was designed to characterize reproducibility challenges in \textit{general} scientific software, predating modern LLM-based workflows. Moreover, it does not cover LLM-specific sources of reproducibility flaws, such as undocumented prompts, stochastic inference behavior, model drift, closed-source models and APIs, missing temperature/seed settings, or the unavailability of training data and model weights~\cite{sallou2023breaking}. To adapt their conceptual structure to the LLM-for-SE context, we replaced their \textit{Human-Centric Issues} category with a new category (\category{Model Smells}) because human factors could not be reliably inferred from our corpus and because LLM-driven research introduces fundamentally new reproducibility risks tied to model behavior rather than researcher actions. Furthermore, we refined the definitions of remaining categories (\eg documentation-, data-, and tool-centric smells) to reflect LLM-specific artifacts such as prompt templates, fine-tuning configurations, API parameters, and model sourcing.

Overall, our taxonomy extends Hassan \etal's foundation by translating a domain-agnostic reproducibility debt categories into an LLM-aware, SE-specific taxonomy that captures emerging reproducibility pitfalls unique to LLM-for-SE research. In doing so, we provide the first systematic, domain-tailored characterization of reproducibility smells for the rapidly evolving LLM-for-SE landscape. In the next subsection, we elaborate on this taxonomy and the categories it contains.


\subsubsection{A Taxonomy of Reproducibility Smells}
Our taxonomy has seven categories, each of which is described below, along with the corresponding detection approach used to identify the smell in papers.

\paragraph{Code and Execution Smells.}
 This type of smell occurs when the paper's experimental artifacts lack a runnable, complete, or accessible implementation needed to replicate results \cite{HASSAN2025112327}. This includes missing code repositories, private or inactive links, incomplete scripts (\eg missing entry points), or a lack of runnable pipelines. These smells reduce \textit{construct validity} by preventing external verification of claims and are exacerbated when models or APIs are closed-source or evolve over time~\cite{Sallou2024}.
 \begin{itemize}
     \item \textit{Identifying code and execution smells}:
For this category, 
we first check whether they provided any link to the source code. If no link was present, we flagged the paper as containing this smell.
If the source code is available, then we checked whether the repository is private, closed-source, or inactive, which also contributed to this category. If the reproducibility links were accessible, we specifically looked for training or inference scripts, shell commands, or Makefiles that could allow an independent researcher to replicate the experiments described in the paper. We flagged the paper as having this smell when the replication package was non-executable (\eg missing entry points), incomplete (\eg missing core scripts, configuration files, or dependency specifications), or entirely absent.  This information was collected by manually opening each repository and inspecting directory structures and execution instructions.
\end{itemize}

\paragraph{Data Smells.}

A \category{data smell} arises when the paper poorly documents the datasets it used or when these used datasets are unavailable, proprietary, or lack stable references such as DOIs, dataset URLs, or version numbers \cite{HASSAN2025112327}. Missing preprocessing details, vague dataset descriptions, or reliance on evaluation data with potential leakage from pretraining exacerbate this issue~\cite{Sallou2024}.
 \begin{itemize}
     \item \textit{Identifying data smells}:
To detect data-related reproducibility issues, we examined whether the papers provided clear references to the datasets used, including persistent identifiers (\eg DOIs, dataset URLs), version numbers, or repository links. We also checked if the dataset was publicly accessible or proprietary. When preprocessing steps were mentioned, we recorded whether they were fully documented or only described vaguely. Missing references, inaccessible datasets, or a lack of preprocessing instructions were labeled as data smells. We systematically extracted this information from both the paper's main text and the linked artifacts, prioritizing README files, data preparation scripts, and supplementary materials.
\end{itemize}
\paragraph{Documentation Smells.}
A \category{documentation smell} is present when methodological descriptions or replication instructions are insufficient, vague, or fragmented. This includes missing or incomplete README files, lack of usage examples, or absence of clear guidance for replicating experiments \cite{HASSAN2025112327}.

 \begin{itemize}
     \item \textit{Identifying documentation smells}: These smels were identified by analyzing both the paper’s methodology section and its associated documentation (\eg README files, Wiki pages, or tutorials). We examined whether details about the file structure, execution steps, or dependencies were missing. We labeled a documentation smell if methodological descriptions were vague, incomplete, or scattered across multiple places without clear guidance for replication.
\end{itemize}
\paragraph{Environment and Tooling Smells.}
An \category{environment and tooling smell} occurs when system dependencies, software environments, or hardware requirements are missing, incomplete, or ambiguous \cite{HASSAN2025112327}. This includes lack of \texttt{requirements.txt}, \texttt{environment.yml}, or \texttt{Dockerfile}, unspecified CUDA or Python versions, or absence of containerization information.
 \begin{itemize}
     \item \textit{Identifying environment and tooling smells}: 
To detect environment and tooling smells, we investigated whether the papers or repositories specified their software and hardware dependencies. We checked for the presence of \texttt{requirements.txt}, \texttt{environment.yml}, or \texttt{Dockerfile} files, and whether specific CUDA, Python, or library versions \etc~were documented. Absence of dependency files, incomplete environment configurations, or missing information about containerization were considered environment smells. Additionally, we noted whether hardware configurations (e.g., GPU type) were mentioned, as a lack of this information can make study replication difficult.
\end{itemize}

\paragraph{Versioning Smells.}
 A \category{versioning smell} is detected when datasets, models, or software dependencies are not associated with explicit version identifiers \cite{HASSAN2025112327}. This includes unpinned package versions, untagged commits, use of floating versions like “latest,” or absent version metadata for models. Lack of versioning severely impacts reproducibility, especially in rapidly evolving LLM environments where model behavior can drift over time~\cite{Sallou2024}. 

 \begin{itemize}
     \item \textit{Identifying versioning smells}:
We classified a reproducibility smell under versioning if datasets, models, or software dependencies lacked clear version information. This included untagged dataset references, unpinned package versions, or ambiguous model identifiers (\eg latest rather than a specific commit or release). We checked the repository, dependency files, and paper text to determine whether versioning was handled rigorously. When version control was missing or inconsistent, we marked it as a versioning smell. It is worth mentioning that when a dataset was used without specifying its version, we tagged the paper with both a data smell (due to insufficient referencing or documentation) and a versioning smell (due to the absence of explicit versioning or stable identifiers).
\end{itemize}

\paragraph{Model Smells.}
As our focus is on using LLM in software engineering, we introduced the category of model smells to capture issues unique to LLM research. 
 A \category{model smell} captures a lack of access to model weights or checkpoints, missing prompt templates, missing details on inference parameters (e.g., temperature, top-k, top-p), or missing fine-tuning hyperparameters. 

 \begin{itemize}
     \item \textit{Identifying model smells}:
We checked whether model weights or checkpoints were available, whether prompt templates were provided, and whether sufficient information about inference parameters (\eg temperature, top-k, top-p) and fine-tuning parameters (\eg learning rate, number of epochs, optimizer) was included. If models were inaccessible, not released, or only vaguely described, the paper was labeled as having a model smell. This information was extracted mainly from the paper. If they were not mentioned, we checked the repository.
\end{itemize}
\paragraph{Access and Legal Smells.}
 An \category{access and legal smell} occurs when experimental artifacts (code, datasets, models) are restricted by licenses, behind approval walls, or have ambiguous usage terms \cite{HASSAN2025112327}. This includes proprietary APIs, restrictive model licenses, institutional access barriers, or a lack of clear licensing statements. 
 
 \begin{itemize}
     \item \textit{Identifying access and legal smells}
We investigated legal and accessibility barriers by checking whether the artifacts (datasets, models, code) were subject to restrictive licenses or hosted behind access requests. We recorded whether licensing information was explicitly stated, whether usage was restricted to specific institutions, or whether artifacts required approval for access. Papers that relied on proprietary tools, private models, or unclear licensing terms were labeled as having access and legal smells. We systematically extracted this data from license files, model cards, and dataset hosting pages.
\end{itemize}

\subsection{Answering RQ1 \& RQ2: Taxonomy-Guided Paper Analysis}
Each paper was manually coded according to the taxonomy, and smell frequencies were aggregated to identify the most prevalent reproducibility barriers in current LLM-for-SE research to answer \textbf{RQ1}. We further analyzed the temporal evolution of these smells on a yearly basis to answer \textbf{RQ2}, enabling us to observe how specific reproducibility barriers have changed over time. This manual analysis was performed by the first and last authors, both of whom have extensive experience serving on artifact evaluation committees at premier software engineering venues. 

\subsection{Answering RQ3: Artifact Badge and Reproducibility Assessment}

To answer \textbf{RQ3}, we systematically examined whether artifact evaluation badges are indicative of stronger reproducibility practices in LLM-for-SE research. Many premier software engineering venues (\eg~ICSE, FSE, ASE) award artifact badges in accordance with ACM or IEEE standards. For example, the ACM’s reproducibility badge system~\cite{acm_badges2020} has the following badges: \ie~\textsf{Artifact Available}, \textsf{Artifact Evaluated–Functional}, \textsf{Artifact Evaluated–Reusable}, and \textsf{Results Reproduced}. IEEE follows a nearly similar badging system \cite{niso_badging,IEEE_Xplore_about_content}, including badges such as \textsf{Open Research Objects (ORO) (Available)}, \textsf{Research Objects Reviewed (ROR)}, \textsf{Results Reproduced (ROR-R)}, and \textsf{Results Replicated (RER)}. These badges are intended to signal robustness and accessibility of research artifacts. 

For each of the 640 papers in our study, we manually inspected official conference proceedings websites, artifact evaluation pages, and paper web pages to record whether an artifact evaluation badge was awarded. When applicable, we also recorded the specific badge type (\eg~\textsf{Available}, \textsf{Functional}, \textsf{Reusable}).

We mapped badge information to reproducibility smells to compare their distributions across all categories (\eg~code and execution, environment and tooling, model, versioning, data, documentation, access, and legal). We inspected whether badge-awarded papers exhibit fewer or less severe smells, and whether particular smell categories are still present despite artifact recognition.

\subsection{Answering RQ4: Publication Venues}
We examine reproducibility practices in the call for papers to understand how venues are enforcing (or not) replication artifacts in published LLM-for-SE works. 

\subsubsection{Selecting Publication Venues for Reproducibility Analysis}
To ensure broad and representative coverage of the LLM-for-SE research landscape, we selected a diverse set of publication venues spanning top-tier software engineering, machine learning, and natural language processing communities to answer RQ4. The venue selection was guided by three key principles: \textbf{(i)} inclusion of \textbf{high-impact venues} based on the CORE ranking system, \textbf{(ii)} representation of both \textbf{conference and journal publications}, and \textbf{(iii)} inclusion of \textbf{preprint platform} to capture rapidly evolving research.

We categorized venues into three main groups: SE venues, ML \& NLP Venues, and pre-print platforms. 

\paragraph{Software Engineering Venues} Core SE publication venues were prioritized, including both A and A*-ranked conferences and journals. These include: 
    \begin{itemize}[leftmargin=15pt]
    \item \textit{\textbf{Journals:}} 
    ACM Transactions on Software Engineering and Methodology (\textbf{TOSEM}), 
    IEEE Transactions on Software Engineering (\textbf{TSE}), 
    Journal of Systems and Software (\textbf{JSS}), 
    Empirical Software Engineering (\textbf{EMSE}), 
    Information and Software Technology (\textbf{IST}).
    
    \item \textit{\textbf{Conferences:}} 
    International Conference on Software Engineering (\textbf{ICSE}), 
    International Conference on Automated Software Engineering (\textbf{ASE}), 
    International Conference on the Foundations of Software Engineering (\textbf{FSE}), 
    IEEE International Conference on Software Maintenance and Evolution (\textbf{ICSME}), 
    International Symposium on Software Testing and Analysis (\textbf{ISSTA}), 
    International Conference on Software Analysis, Evolution and Reengineering (\textbf{SANER}), 
    International Conference on Program Comprehension (\textbf{ICPC}), 
    Mining Software Repositories Conference (\textbf{MSR}), 
    International Symposium on Empirical Software Engineering and Measurement (\textbf{ESEM}), 
    International Conference on Evaluation and Assessment in Software Engineering (\textbf{EASE}), 
    IEEE International Conference on Software Testing, Verification and Validation (\textbf{ICST}), 
    IEEE International Symposium on Software Reliability Engineering (\textbf{ISSRE}), 
    IEEE International Conference on Software Architecture (\textbf{ICSA}).
\end{itemize}
    These venues represent the main outlets for software engineering research and are frequently used for reporting empirical studies, tool evaluations, and engineering-focused LLM applications.

\paragraph{ML and NLP Venues} To capture research at the intersection of LLMs and SE, we also included leading ML and NLP conferences:
    \begin{itemize}[leftmargin=15pt]
    \item \textit{\textbf{Conferences:}} 
    Conference on Neural Information Processing Systems (\textbf{NeurIPS}), 
    International Conference on Machine Learning (\textbf{ICML}), 
    Conference on Empirical Methods in Natural Language Processing (\textbf{EMNLP}), 
    AAAI Conference on Artificial Intelligence (\textbf{AAAI}), 
    International Conference on Learning Representations (\textbf{ICLR}), 
    Conference on Learning Theory (\textbf{COLT}), 
    International Joint Conference on Artificial Intelligence (\textbf{IJCAI}), 
    International Conference on Principles of Knowledge Representation and Reasoning (\textbf{KR}), 
    Annual Meeting of the Association for Computational Linguistics (\textbf{ACL}).
\end{itemize}
    These venues often publish foundational work on LLM architectures, training methods, prompt engineering, and model evaluation that can be applied to SE tasks.

\paragraph{Pre-print Platforms} Given the rapid pace of development in LLM research, many influential works appear first as pre-prints before formal peer review. We therefore included \textbf{arXiv} as an additional source of recent and emerging studies.

\subsubsection{Reproducibility Practices Analysis}
To answer RQ4, we systematically analyzed the aforementioned publication venues' calls for papers and websites from 2020 to 2025 (\ie the same year range as our 640 studied papers) to understand how institutional policies and review processes shape reproducibility practices in LLM-for-SE research. For each venue, we checked whether it provides a formal artifact evaluation track or equivalent mechanism, supports or requires reproducibility badges (\eg~\textit{artifact available}, \textit{artifact evaluated}, \textit{results reproduced}), mandates code or dataset availability as part of the publication process, and includes explicit review guidelines or checklists addressing reproducibility and transparency.
\section{Results}
\label{sec:results}

This section presents the findings of our study, highlighting the prevalence, characteristics, and temporal evolution of reproducibility smells in LLM-for-SE research. Furthermore, we discuss artifact evaluation badges' reliability in terms of reproducibility smells as well as the reproducibility practices for publication venues.

\subsection{RQ1: Prevalence of Reproducibility Smells}

\subsubsection{Reproducibility Smells Distribution}
Table \ref{tab:rq1_smell_distribution} summarizes the overall distribution of reproducibility smells across all the \textbf{640} analyzed LLM-for-SE papers. From our analysis, we found that  \category{access and legal smells} (\textbf{35.9\%}) and \category{code and execution smells} (\textbf{35.5\%}) are the most frequent issues, indicating that studies either lack runnable artifacts or impose legal and licensing barriers that complicate replication. \category{Versioning smells} are also widespread (\textbf{32.2\%}), reflecting common problems such as unpinned dependency versions, missing model or dataset identifiers, and ambiguous versioning practices.




\begin{table}[!htbp]
    \footnotesize
    \centering
    \caption{Reproducibility Smells Distribution.}
\label{tab:rq1_smell_distribution}
    \begin{tabular}{ c   c   c   }
    \toprule
\textbf{Reproducibility Smell} & \textbf{\# Papers} & \textbf{\% Papers} \\
         \midrule
Access and Legal & 230 & 35.9\% \\
Code and Execution & 227 & 35.5\% \\
Versioning & 206 & 32.2\% \\
Environment and Tooling & 135 & 21.1\% \\
Model & 59 & 9.2\% \\
Data & 51 & 8.0\% \\
Documentation & 23 & 3.6\% \\
\midrule
\textit{No Smells} & 85 & 13.3\% \\
\bottomrule
    \end{tabular}
    
\end{table}

\category{Environment and tooling smells} appear in 21.1\% of papers, pointing to gaps in specifying computational dependencies, environment configuration files, or hardware details. In contrast, \category{data} (\textbf{8\%}) and \category{model smells} (\textbf{9.2\%}) occur less frequently, but they still signal gaps in dataset accessibility, model checkpoint availability, or missing prompt and inference parameter details. \category{Documentation smells} are the least pervasive category (\textbf{3.6\%}), suggesting that while methodological descriptions are often present, other factors, such as execution or access barriers, remain more dominant. Finally, only \textbf{13.3\%} of papers did not have any reproducibility smells, indicating that reproducibility challenges are widespread in LLM-for-SE literature. These findings show that while LLM-for-SE research often provides some documentation, there remain open challenges in reproducing these works related to executable artifacts, access constraints, and transparency in versioning.

\subsubsection{Reproducibility Smells Co-occurrence}
\label{subsec:structural_entanglement}

\begin{figure}[!htbp]
    \centering
    \includegraphics[width=\linewidth]{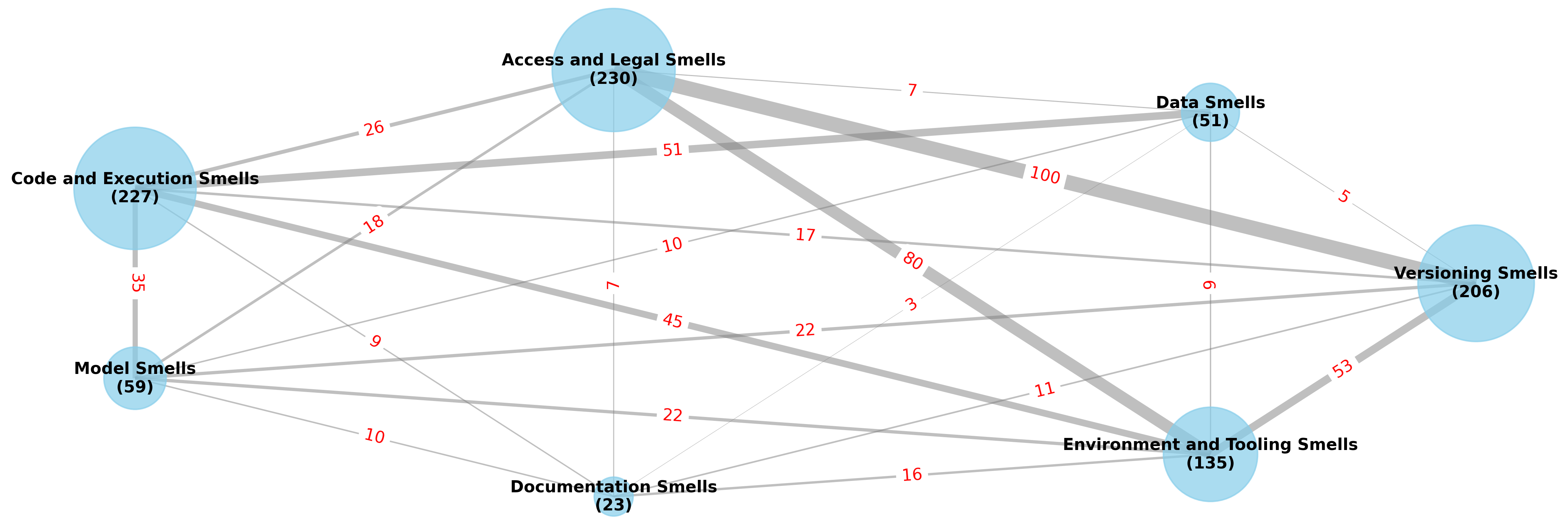}
    \caption{Co-occurrence Network of Reproducibility Smells.}
    \label{fig:occurrence}
\end{figure}

Our co-occurrence analysis of the reproducibility smells reveals a structural coupling among several smell categories, as shown in Figure \ref{fig:occurrence}, indicating that reproducibility debts are often interconnected. For example, among 230 papers with \category{Access and Legal} smells, \textbf{43.5\%} (100 papers) also exhibit \category{Versioning} smells. This reflects a common pattern in LLM-for-SE work, where restricted or proprietary APIs coincide with undocumented or unstable software versions, making it more difficult to reproduce experiments.

A similar co-occurrence is observed between the \category{Access and Legal} and \category{Environment and Tooling} smells. \textbf{34.8\%} (80 of 230) of papers with access-related issues also lack sufficient environment specifications, suggesting that institutional barriers (\eg licensing, rate limits, or unavailable datasets) propagate into incomplete environment descriptions in practice.

\category{Code and Execution} smells also demonstrate several meaningful co-occurrences. For example, \textbf{22.5\%} (51 of 227) of papers with code issues also suffer from \category{data smells}, while \textbf{19.8\%} (45 of 227) co-occur with environment and tooling issues. These results show  an interplay between runnable pipelines, data availability, and environment specification—three ingredients essential for end-to-end reproducibility.

To quantify the strength of these relationships, we computed pairwise lift (\ie measures how much more frequently two smells co-occur than would be expected if they were independent) and $\phi$-coefficients (\ie captures the correlation strength between two binary smell variables) for all smell combinations. The strongest structural associations were observed between \category{Documentation} and \category{Model} smells (lift $\approx 4.72$, $\phi \approx 0.23$) and between \category{Documentation} and \category{Environment and Tooling} smells (lift $\approx 3.30$, $\phi \approx 0.23$), followed by \category{Code and Execution} and \category{Data} smells (lift $\approx 2.82$, $\phi \approx 0.40$), indicating \textbf{pronounced}, \textbf{non-random co-dependencies.} We also find elevated associations among infrastructure-related smells, specifically, \category{Access and Legal}, \category{Environment and Tooling}, and \category{Versioning} (lift up to $\approx 1.65$, $\phi$ up to $\approx 0.25$). This reinforces that reproducibility debt is organized into coherent clusters rather than isolated issues.


These patterns demonstrate that reproducibility debt in LLM-for-SE research is \emph{structural}. If we make improvements to any single artifact dimension (\eg documentation), we may not fully resolve the problem unless we address other issues like access constraints, code executability, dependency management, and model availability.

\subsection{RQ2: Temporal Trends in Reproducibility Over Time}

\begin{figure}[!htbp]
    \centering
    \includegraphics[width=\linewidth]{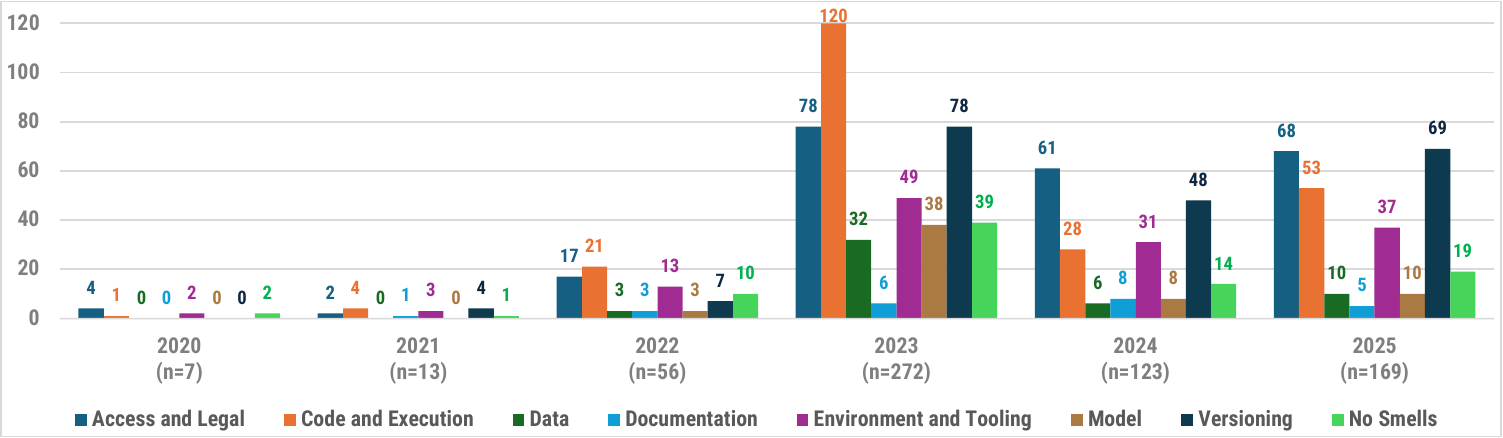}
   \caption{Temporal distribution of reproducibility smells across years. The total number of papers with reproducibility smells may exceed the total number of papers, as multiple smells can co-occur within a single paper.}
\label{fig:rq2_smell_trends}
\end{figure}

Figure~\ref{fig:rq2_smell_trends} presents the temporal distribution of reproducibility smells across publication years from 2020 to 2025. 
%
%
Across the analyzed period, we found that \category{code and execution smells} remain the most prevalent category overall. In 2020, only 1 out of 7 papers that year had this smell (14.3\%), rising to 120 papers in 2023 (44.1\% of papers in that year), before declining to 28 papers (22.8\%) in 2024 and 53 papers (31.4\%) in 2025. This downward trend in recent years suggests incremental improvements in the availability and clarity of experimental pipelines. A chi-square test of independence confirms that the distribution of reproducibility smells varies significantly across years ($\chi^2=75.34$, $p=0.000089<0.001$), providing statistical support for the temporal trends observed in Figure~\ref{fig:rq2_smell_trends}.

\category{Access and legal smells} show an increase over time. Starting at 4 papers (57.1\%) in 2020, their prevalence rises to 61 papers (49.6\%) in 2024 and 68 papers (40.2\%) in 2025, making this category one of the frequent reproducibility barriers in recent research. \category{Versioning smells} follow a similar pattern, increasing from 0 papers in 2020 to 48 (39.0\%) in 2024 and 69 (40.8\%) in 2025. These two categories show the sharpest rise over the six-year span we analyzed, reflecting growing challenges related to model version pinning, licensing, and access restrictions.

\category{Environment and tooling smells} appear consistently across the years. There were 2 papers (28.6\%) in 2020, 49 (18.0\%) in 2023, and 37 (21.9\%) in 2025, indicating persistent, unresolved issues in dependency specification and computational reproducibility. 
Moreover, \category{data smells} and \category{documentation smells} are relatively infrequent. Data smells peak at 32 papers (11.8\%) in 2023 but remain below 6\% in other years. Documentation smells never exceed 8 papers in any year, with percentages ranging from 0.0\% to 7.7\%, suggesting that these issues, while present, are less widespread compared to other categories.

\category{Model smells} emerge more prominently in and after 2022. There were 38 papers (14.0\%) in 2023 and 10 papers (5.9\%) in 2025 with these smells, highlighting growing dependencies on closed-source checkpoints or unreleased model components.
Finally, the proportion of papers with \textbf{\textit{no detected reproducibility smells}} declines slightly over time: from 28.6\% in 2020 to 11.2\% in 2025. This indicates that as the LLM-for-SE research landscape expands, reproducibility issues remain pervasive, particularly those related to legal, versioning, and environment concerns.

\subsection{RQ3: Artifact Evaluation Badges Reliability}
Table~\ref{tab:rq_badges_smells_trends} and Table~\ref{tab:rq_badges_smells_mapping} summarize the evolution of artifact evaluation badges and reproducibility smells from \textbf{2020} to \textbf{2025}. We found 63 papers (9.8\%) with badges. It is worth mentioning that papers from 2024 and 2025 in our dataset primarily originate from the top three software engineering venues (ICSE, FSE, ASE), whereas earlier years include a broader mix of journals, conferences, and pre-prints, and the number of badged papers in our corpus is small (especially before 2023). For this reason, our badge analysis is intentionally associational rather than causal. We stratify the results descriptively by year and badge combination (Tables~\ref{tab:rq_badges_smells_trends} and~\ref{tab:rq_badges_smells_mapping}), and we interpret badges as \emph{proxies} for stronger artifact scrutiny rather than as direct causal interventions.

It is also worth noting that only two papers used the IEEE badging system \cite{TOSEM-LLM4SE_218, TOSEM-LLM4SE_386} and had only the \textsf{Available} badge. Two other papers used adapted ACM badging systems with similar badge categories \cite{TOSEM-LLM4SE_57,TOSEM-LLM4SE_312}. For this reason, Table~\ref{tab:rq_badges_smells_trends} and Table~\ref{tab:rq_badges_smells_mapping} list ACM-style badging categories \ie~\textsf{Artifact Available}, \textsf{Artifact Evaluated—Functional}, and \textsf{Artifact Evaluated—Reusable}. \textsf{Artifact Available} indicates that the authors have uploaded their artifacts (\eg~code, data, models) in a stable, publicly accessible repository, while \textsf{Artifact Evaluated—Functional} certifies that reviewers were able to execute the artifact and verify its basic correctness. \textsf{Artifact Evaluated—Reusable} reflects a higher standard, recognizing artifacts that are not only functional but also well-documented, modular, and easy for others to adapt or extend.

\begin{table}[!htbp]
\centering
\footnotesize
\caption{Temporal distribution of badges.}

\label{tab:rq_badges_smells_trends}
\setlength{\tabcolsep}{6pt} 
{\begin{tabular}{@{}lcccccccccccc@{}}
\toprule
\multirow{2}{*}{\textbf{Category}}  & \multicolumn{2}{c}{\begin{tabular}[c]{@{}c@{}}\textbf{2020}\\\textbf{(n=7)}\end{tabular}} & \multicolumn{2}{c}{\begin{tabular}[c]{@{}c@{}}\textbf{2021}\\\textbf{(n=13)}\end{tabular}} & \multicolumn{2}{c}{\begin{tabular}[c]{@{}c@{}}\textbf{2022}\\\textbf{(n=56)}\end{tabular}} & \multicolumn{2}{c}{\begin{tabular}[c]{@{}c@{}}\textbf{2023}\\\textbf{(n=272)}\end{tabular}} & \multicolumn{2}{c}{\begin{tabular}[c]{@{}c@{}}\textbf{2024}\\\textbf{(n=123)}\end{tabular}} & \multicolumn{2}{c}{\begin{tabular}[c]{@{}c@{}}\textbf{2025}\\\textbf{(n=169)}\end{tabular}} \\
\\
 \cmidrule(lr){2-3} \cmidrule(lr){4-5} \cmidrule(lr){6-7} \cmidrule(lr){8-9} \cmidrule(lr){10-11} \cmidrule(lr){12-13}
 & \textbf{\#} & \textbf{\%} & \textbf{\#} & \textbf{\%} & \textbf{\#} & \textbf{\%} & \textbf{\#} & \textbf{\%} & \textbf{\#} & \textbf{\%} & \textbf{\#} & \textbf{\%} \\
\midrule
At least one badge & 1 & 14.3 & 1 & 7.7 & 4 & 7.1 & 11 & 4.0 & 12 & 9.8 & 34 & 20.1 \\ \hline
Available & 1 & 14.3 & 1 & 7.7 & 3 & 5.4 & 11 & 4.0 & 12 & 9.8 & 34 & 20.1 \\
Functional & 0 & 0.0 & 1 & 7.7 & 2 & 3.6 & 5 & 1.8 & 2 & 1.6 & 22 & 13.0 \\
Reusable & 1 & 14.3 & 0 & 0.0 & 0 & 0.0 & 1 & 0.4 & 7 & 5.7 & 17 & 10.1 \\
\bottomrule
\end{tabular}}
\end{table}


\begin{table}[!htbp]
\centering
\footnotesize
\caption{Mapping between badge combinations and reproducibility smells in badged papers from 2020 -- 2025. Each row represents papers sharing the same year and badge combination (\textbf{A} means available, \textbf{F} means functional, \textbf{R} means Reusable). Each cell shows the count of papers exhibiting each smell, followed by the percentage relative to the group size.}
\label{tab:rq_badges_smells_mapping}
\setlength{\tabcolsep}{6pt} 
\begin{tabular}{@{}ccrccccccc@{}}
\toprule
\textbf{Year} & \textbf{Badge} & \textbf{\#} & \textbf{Legal} & \textbf{Code} & \textbf{Data} & \textbf{Doc} & \textbf{Env} & \textbf{Model} & \textbf{Version} \\
\midrule
2020 & A+R & 1 & 0 (0.0) & 1 (100.0) & 0 (0.0) & 0 (0.0) & 1 (100.0) & 0 (0.0) & 0 (0.0) \\\hline
2021 & A+F & 1 & 0 (0.0) & 0 (0.0) & 0 (0.0) & 0 (0.0) & 0 (0.0) & 0 (0.0) & 1 (100.0) \\\hline
2022 & A & 2 & 1 (50.0) & 0 (0.0) & 0 (0.0) & 0 (0.0) & 0 (0.0) & 0 (0.0) & 1 (50.0) \\
2022 & A+F & 1 & 0 (0.0) & 1 (100.0) & 0 (0.0) & 0 (0.0) & 0 (0.0) & 0 (0.0) & 0 (0.0) \\
2022 & F & 1 & 0 (0.0) & 1 (100.0) & 0 (0.0) & 0 (0.0) & 0 (0.0) & 0 (0.0) & 0 (0.0) \\\hline
2023 & A & 6 & 0 (0.0) & 2 (33.3) & 0 (0.0) & 0 (0.0) & 0 (0.0) & 0 (0.0) & 1 (16.7) \\
2023 & A+F & 4 & 1 (25.0) & 1 (25.0) & 1 (25.0) & 0 (0.0) & 2 (50.0) & 1 (25.0) & 1 (25.0) \\
2023 & A+F+R & 1 & 0 (0.0) & 0 (0.0) & 0 (0.0) & 0 (0.0) & 0 (0.0) & 0 (0.0) & 1 (100.0) \\\hline
2024 & A & 3 & 0 (0.0) & 0 (0.0) & 0 (0.0) & 1 (33.3) & 1 (33.3) & 0 (0.0) & 2 (66.7) \\
2024 & A+F & 2 & 1 (50.0) & 0 (0.0) & 0 (0.0) & 0 (0.0) & 0 (0.0) & 0 (0.0) & 0 (0.0) \\
2024 & A+R & 7 & 2 (28.6) & 0 (0.0) & 0 (0.0) & 0 (0.0) & 0 (0.0) & 0 (0.0) & 3 (42.9) \\\hline
2025 & A & 12 & 2 (16.7) & 1 (8.3) & 0 (0.0) & 0 (0.0) & 2 (16.7) & 1 (8.3) & 6 (50.0) \\
2025 & A+F & 5 & 2 (40.0) & 1 (20.0) & 0 (0.0) & 0 (0.0) & 0 (0.0) & 0 (0.0) & 3 (60.0) \\
2025 & A+F+R & 17 & 2 (11.8) & 5 (29.4) & 1 (5.9) & 1 (5.9) & 2 (11.8) & 2 (11.8) & 8 (47.1) \\
\bottomrule
\end{tabular}
\end{table}

Badge adoption remained sparse until recently: fewer than \textbf{8\%} of papers prior to 2023 earned badges, increasing to \textbf{9.8\%} in 2024 and \textbf{20.1\%} in 2025. This uptick aligns with stronger enforcement and visibility of artifact evaluation processes at flagship SE venues. The \textsf{Artifact Available} badge was consistently the most common, while \textsf{Functional} and \textsf{Reusable} badges were less frequent until 2025, where both incidences rose to \textbf{13.0\%} and \textbf{10.1\%}, respectively. This upward trend reflects growing expectations not only for availability but also for usability and reusability of research artifacts.

Despite these positive developments, reproducibility smells persist even among badged papers. Across years, the most common smells among badge recipients were \category{versioning issues} (reaching \textbf{50\%} in 2025) and \category{code and execution smells}. Notably, some earlier badged papers (with very small sample sizes) exhibited high smell incidence rates, highlighting that early badges may have focused primarily on artifact presence rather than quality or usability. Even in 2024–2025, when SE venues set higher reproducibility standards, badged papers frequently lacked explicit dependency versions or stable execution pipelines. For example, \textbf{41.7\%} of badged papers from 2024 still exhibited \category{versioning smells}.

\subsection{RQ4: Reproducibility Practices Across Publication Venues}
To address RQ4, we analyzed reproducibility practices across 28 publication venues spanning the software engineering (SE), machine learning (ML), and natural language processing (NLP) communities. Our analysis focused on four key dimensions: \textbf{(i)} presence of artifact evaluation tracks, \textbf{(ii)} support for reproducibility badges, \textbf{(iii)} policies requiring or encouraging sharing of code and datasets, and \textbf{(iv)} use of formal review checklists addressing reproducibility.

\begin{table*}[!htbp]
\centering
\footnotesize
\caption{Reproducibility support across publication venues based on the latest available data on November 2025. Full black circle = present/required, half black circle = partial/encouraged, empty black circle = absent}
\label{tab:venue_reproducibility_summary}
\setlength{\tabcolsep}{6pt}
\begin{tabular}{l c c c c}
\toprule
\textbf{Venue} & \textbf{Artifact Evaluation Track} & \textbf{Badges} & \textbf{Code/Data} & \textbf{Checklist} \\
\midrule
\multicolumn{5}{l}{\textit{\textbf{SE Journals}}} \\
TOSEM & \full & \full & \full & \none \\
EMSE  & \none & \none & \full & \none \\
JSS   & \none & \none & \full & \none \\
IST   & \none & \none & \full & \none \\
TSE   & \none & \none & \half & \none \\
\midrule
\multicolumn{5}{l}{\textit{\textbf{SE Conferences (ACM)}}} \\
ICSE  & \full & \full & \full & \none \\
FSE   & \full & \full & \full & \none \\
ASE   & \full & \full & \full & \none \\
ISSTA & \full & \full & \full & \none \\
\midrule
\multicolumn{5}{l}{\textit{\textbf{SE Conferences (IEEE)}}} \\
ICSME & \full & \full & \full & \none \\
ISSRE & \full & \full & \full & \none \\
ICSA  & \full & \full & \full & \none \\
\midrule
\multicolumn{5}{l}{\textit{\textbf{Other SE Conferences}}} \\
SANER & \half & \none & \full & \none \\
MSR   & \none & \none & \full & \none \\
ESEM  & \none & \none & \full & \none \\
ICPC  & \none & \none & \full & \none \\
ICST  & \none & \none & \half & \none \\
EASE  & \none & \none & \half & \none \\
\midrule
\multicolumn{5}{l}{\textit{\textbf{ML Conferences}}} \\
NeurIPS & \none & \none & \full & \full \\
ICML   & \none & \none & \full & \full \\
ICLR   & \none & \none & \full & \full \\
AAAI   & \none & \none & \full & \full \\
IJCAI  & \none & \none & \full & \full \\
COLT   & \none & \none & \none & \none \\
KR     & \none & \none & \half & \none \\
\midrule
\multicolumn{5}{l}{\textit{\textbf{NLP Conferences}}} \\
ACL    & \none & \none & \full & \full \\
EMNLP  & \none & \none & \full & \full \\
\midrule
\multicolumn{5}{l}{\textit{\textbf{Preprints}}} \\
arXiv & \none & \none & \half & \none \\
\bottomrule
\end{tabular}
\end{table*}

\begin{itemize}[leftmargin=*,label=--]
    \item \textbf{SE Journals.} Among the SE journals, \textsf{TOSEM} offers the most structured support, providing optional artifact submission via Replicated Computational Results (RCR) reports and endorsing ACM badges. However, this was introduced in 2022 and implemented from 2023 onwards \cite{tosem2022why}. \textsf{EMSE} actively promotes reproducibility through its Open Science initiative, requiring data availability statements and encouraging artifact sharing with informal review by its Open Science Board from 2019 \cite{emse_openscience}. \textsf{JSS} also promotes reproducibility through the JSS Open Science Initiative \cite{jss_openscience_guidelines}. For papers where authors indicate participation in the initiative, the manuscript is automatically forwarded to the JSS Open Science Board after acceptance. Following a successful review of availability and usability, the publisher will add a statement acknowledging validation of the Open Science material. \textsf{IST} encourages artifact sharing and requires data availability statements, though they lack formal evaluation or badging \cite{ist_guide_for_authors}. \textsf{TSE}, as an IEEE journal, follows IEEE's general policy of encouraging (but not requiring) artifact availability and maintains this policy from 2020 to 2025.

\item \textbf{SE Conferences.} ACM-affiliated SE conferences such as \textsf{ICSE}, \textsf{FSE}, \textsf{ASE}, and \textsf{ISSTA} feature dedicated artifact evaluation tracks throughout 2020 to 2025. These venues apply ACM's reproducibility badge system, introduced in 2016 and updated to version 1.1 in 2020 \cite{acm_badges2020}. This badging system includes the following badge categories: \textit{Artifact Available}, \textit{Artifact Evaluated – Reusable}, and \textit {Results Reproduced} to qualifying papers. IEEE venues, including \textsf{ICSME}, \textsf{ISSRE}, and \textsf{ICSA}, have adopted similar practices, offering artifact tracks and reproducibility badges based on the IEEE-endorsed \textit{Research Object Reviewed (ROR)}/\textit{Open Research Object (ORO)} framework. However, \textsf{ICSME} introduced artifact badging in pilot mode in 2020, awarding ACM badges \cite{icsme2020_artifacts}, and adopted the current IEEE-endorsed badging scheme starting in 2021. Similarly, \textsf{ISSRE} initiated its artifact badging process in 2023. ICSA launched its first Artifact Evaluation track in 2021 and expanded it in 2022 \cite{icsa2022_artifact}. The ICSA process is built on the National Information Standards Organization (NISO) reproducibility badge definitions (supported by IEEE) and mirrors ACM’s criteria \cite{niso_badging}.

Other SE conferences, such as \textsf{SANER}, \textsf{MSR}, \textsf{ESEM}, and \textsf{ICPC}, do not employ formal artifact evaluation but enforce strong open science policies that require or strongly encourage code/data sharing and mandate data availability statements at the time of paper submission. Moreover, \textsf{SANER} hosts specialized tracks introduced in 2023 for reproducibility and replication studies, providing structured incentives without formal badging \cite{saner2023_rene}. \textsf{ICST} and \textsf{EASE} promote transparency and artifact sharing informally, without specific evaluation mechanisms.

\item \textbf{ML and NLP Conferences.} Leading ML conferences such as \textsf{NeurIPS}, \textsf{ICML}, and \textsf{ICLR} enforce reproducibility through detailed checklists that authors must complete upon submission. These checklists cover code and data availability, computational details, and reporting standards. Although these venues do not issue badges, reproducibility is integrated into the review criteria, and post-publication reproducibility challenges further incentivize transparency. \textsf{NeurIPS} introduced this reproducibility program in 2019, requiring authors to complete a detailed checklist with each submission. In 2021, NeurIPS expanded it into a broader ``responsible research'' checklist (covering reproducibility, transparency, ethics, and societal impact) \cite{neurips2021_checklist}. In 2020, \textsf{ICML} introduced a similar reproducibility checklist for authors (modeled on NeurIPS’s) \cite{ICML2020StyleAuthorInstructions}. \textsf{ICLR} began urging authors to include a ``Reproducibility Statement'' in their submissions around 2021 and to submit code; by 2025, ICLR’s author guide explicitly stated that it ``strongly encourages'' a reproducibility section in each paper \cite{ICLR2026AuthorGuide}.

\textsf{AAAI} 2021 was the first AAAI conference to require a reproducibility checklist with each submission \cite{Magnusson2023}. This checklist, similar to NeurIPS’s, became mandatory for AAAI submissions. \textsf{IJCAI} has followed a similar suit, requiring checklists and reproducibility assessments by reviewers by explicitly incorporating reproducibility scores into the review form. 

From 2020 to 2025, \textsf{COLT} maintained a strict focus on theoretical reproducibility through complete proofs, without introducing artifact sharing policies, reproducibility checklists, or badges \cite{colt2023cfp}. Experimental components were optional and not subject to formal reproducibility expectations throughout this period. Between 2020 and 2025, \textsf{KR} gradually expanded support for reproducibility by allowing supplementary materials (\eg~proofs, code, or data), and by 2025 explicitly encouraged artifact sharing, particularly in applied tracks. However, it did not adopt artifact evaluation or badging \cite{kr2025cfp, kr2024cfp}.

In the NLP domain, \textsf{EMNLP} and \textsf{ACL} enforce reproducibility through the Responsible NLP Checklist, mandatory limitations sections, and expectations for artifact availability statements in camera-ready submissions from 2020 \cite{Magnusson2023}. While these venues do not operate formal artifact tracks, community norms strongly encourage public release of code and datasets.

\item \textbf{Preprints.} \textsf{arXiv}, as a preprint platform, does not enforce any reproducibility requirements. Artifact availability is left to the discretion of authors, with no peer review or badging mechanisms. There is a separate tab, introduced in 2020 \cite{arXivLabs2020CodeFeature}, in each of the Machine Learning arXiv papers to provide the link to data and code using third-party services like Hugging Face, Papers with Code \etc

\end{itemize}

\section{Recommendations}
\label{sec:reco}
The findings from RQ1–RQ4 show that LLM-for-SE research could benefit from stronger, more systematic reproducibility practices. In this section, we present actionable recommendations for each reproducibility smell category, offering concrete steps that authors, reviewers, and venues could adopt to prevent persistent barriers from becoming entrenched. We also introduce a  \textbf{Reproducibility Maturity Model (RMM)} to move beyond binary notions of artifact quality and provide a structured, multi-dimensional framework for assessing the durability, rigor, and long-term verifiability of research outputs.

\subsection{ Smells Prevalence and Actionable Recommendations}

Our analysis across 640 LLM-for-SE papers reveals that reproducibility challenges remain a problem. The most common problem areas are \category{Access and Legal} and \category{Code and Execution} smells (each affecting roughly one-third of the studies), followed closely by \category{Versioning} issues (32.2\%). \category{Environment and Tooling} smells (21.1\%) is less frequent but notable. Other smells like \category{Data}, \category{Model}, and \category{Documentation} smells occur below 10\%. In the next sections, we discuss each smell category with recommendations.

\subsubsection{Access and Legal Smells}

\category{Access and Legal Smells} is one of most pervasive issues, affecting \textbf{35.9\%} of studies overall, with a sharp rise from \textbf{28.7\%} in 2023 to nearly half of all papers (\textbf{49.6\%}) in 2024. This rise parallels the widespread adoption of closed APIs and gated model endpoints. Common signs of this smell are replication packages with missing or non-permissive licenses and reliance on commercial APIs that cannot be redistributed. The absence of clear licensing and access documentation creates legal uncertainty for reuse and hinders downstream replication.

\paragraph{Recommendations:}
\begin{itemize}
    \item \textbf{Mandatory Licensing:} Every artifact must specify an explicit license (\eg MIT, Apache 2.0, CC-BY-4.0). Artifact Evaluation (AE) forms should be rejected if licensing is absent or ambiguous.
    \item \textbf{Persistent Hosting:} Encourage repositories such as Zenodo or Figshare for long-term archival access to code, data, and model weights beyond GitHub links.
    \item \textbf{Transparent Proprietary Usage:} For closed APIs, papers should clearly document API version, model identifier, and date of access, and provide request/response examples to enable approximate reproduction.
\end{itemize}

\subsubsection{Code and Execution Smells}

\category{Code and Execution Smells} affect \textbf{35.5\%} of papers but show a modest improvement post-2023, dropping from 44.1\% in 2023 to 31.4\% in 2025. However, this issue still persists because inactive or private repositories, missing entry points, and incomplete scripts remain frequent. Out of 477 replication links examined, 19 were broken or private, and 64 contained incomplete or non-runnable artifacts. This confirms that the mere existence of shared code does not guarantee reproducibility.

\paragraph{Recommendations:}
\begin{itemize}
    \item \textbf{Runnable Entry Points:} AE guidelines should require a single executable entry (\eg \texttt{run.sh}, \texttt{Makefile} target, or Jupyter notebook) verified during artifact review.
    \item \textbf{Reproducible Pipelines:} Encourage standardized workflows (Make, Snakemake, or bash scripts) that automate preprocessing, training, and evaluation.
    \item \textbf{Code Completeness:} All scripts tied to reported results must be present and documented within the replication package.
\end{itemize}

\subsubsection{Versioning Smells}

\category{Versioning Smells} were found in \textbf{32.2\%} of papers, rising from 12.5\% in 2022 to over 40\% in 2024–2025. The most common issues include missing dependency specifications, unpinned library versions, and vague references such as ``latest model release''.

\paragraph{Recommendations:}
\begin{itemize}
    \item \textbf{Pin All Dependencies:} Require exact version locking using environmental files (\eg \texttt{requirements.txt} or \texttt{environment.yml}) with explicit equality operators.
    \item \textbf{Model Version Disclosure:} Papers must specify model commit hashes, release tags, or dataset identifiers for all LLMs used.
    \item \textbf{Data Versioning:} Track derived datasets via  Data Version Control (DVC) or Digital Object Identifier (DOI)-based repositories to ensure stable data lineage.
\end{itemize}

\subsubsection{Environment and Tooling Smells}

Roughly \textbf{21.1\%} of studies have \category{Environment and Tooling Smells}, a proportion that has remained stable across years (around 18–25\%). Missing or incompatible environment settings frequently cause replication failures, and only 33 papers include Docker support. Furthermore, \textbf{572 papers} lacked any mention of hardware setup, obscuring critical GPU and driver dependencies.

\paragraph{Recommendations:}
\begin{itemize}
    \item \textbf{Containerization:} Require a runnable \texttt{Dockerfile} or similar container specification to guarantee reproducible environments.
    \item \textbf{Hardware Disclosure:} Explicitly report GPU/CPU models, CUDA versions, and OS details, especially for fine-tuning experiments.
    \item \textbf{Executable Environment Files:} Ensure environment specifications are directly runnable and reflect all transitive dependencies.
\end{itemize}

\subsubsection{Model Smells}

\category{Model Smells} occur in \textbf{9.2\%} of papers and are  tied to ambiguous model configurations. Missing prompt templates, undocumented inference parameters, and absent training details cause this smell. For instance, 199 papers failed to specify their exact prompt templates, and 88 lacked documentation of their inference parameters.

\paragraph{Recommendations:}
\begin{itemize}
    \item \textbf{Release Fine-tuned Weights:} When feasible, publish fine-tuned checkpoints or scripts to regenerate them.
    \item \textbf{Parameter Transparency:} Report all inference and training parameters, including sampling temperature, top-$k$, and learning rates.
    \item \textbf{Prompt Publication:} Archive all prompt templates used in evaluation to ensure consistent task reproduction.
\end{itemize}

\subsubsection{Data Smells}

Detected in \textbf{7.9\%} of studies, \category{Data Smells} typically occur from missing preprocessing documentation or inaccessible dataset versions. Even when datasets were cited in the paper, the artifact often lacked persistent links or transformation scripts.

\paragraph{Recommendations:}
\begin{itemize}
    \item \textbf{Stable Data References:} Use DOIs or permanent URLs for datasets and their processed variants.
    \item \textbf{Include Preprocessing Scripts:} Provide runnable scripts that transform raw data into the experiment-ready format.
    \item \textbf{Dataset Cards:} Summarize data provenance, filtering, and versioning in concise dataset documentation.
\end{itemize}

\subsubsection{Documentation Smells}

Although \category{Documentation Smells} is the least frequent (\textbf{3.6\%}), it can critically block replication when other artifacts are ambiguous. Even when the code is complete, missing README files or disjointed instructions often make it practically unusable.

\paragraph{Recommendations:}
\begin{itemize}
    \item \textbf{Unified README:} Include a single, step-by-step \texttt{README.md} that maps each experiment to paper results.
    \item \textbf{Minimal Documentation Principle:} Centralize all essential instructions in one or two top-level documents.
    \item \textbf{Self-contained Guidance:} Documentation should enable replication without author contact or guesswork.
\end{itemize}

\subsection{Toward a Reproducibility Maturity Model (RMM)}
\label{sec:repro_maturity_model}

As discussed in RQ3 and RQ4, ACM and IEEE provide artifact badges \ie ``available'', ``functional'', or ``reusable'', but they do not offer guidance on the \textit{durability}, \textit{rigor}, or \textit{long-term verifiability} of research outputs. Our empirical analysis shows that these systems primarily evaluate \emph{momentary functionality}. That means they check whether artifacts worked during review, but not whether they are engineered for long-term reproducibility. Over 40\% of “functional” artifacts in our corpus from 2024 -- 2025 fail within months due to drifting dependencies, unpinned versions, incomplete environments, or unclear licensing. Moreover, existing badges do not assess reproducibility holistically across the multiple smell dimensions we identified. A paper can earn a badge even when substantial barriers remain, as shown in RQ3.

To address this gap, we propose a \textbf{Reproducibility Maturity Model (RMM)}, a multi-dimensional framework that evaluates the \emph{durability}, \emph{rigor}, and \emph{legal openness} of artifacts. It is designed to help authors, reviewers, and conference organizers systematically evaluate and enforce the reproducibility of LLM-for-SE research.

The RMM framework evaluates reproducibility along five orthogonal axes identified through our taxonomy and empirical observations:
\begin{enumerate}[label=\textbf{A\arabic*.}, leftmargin=*]
    \item \textbf{Accessibility.} Availability, persistence, and openness of datasets, code, and models.
    \item \textbf{Environment Specification.} Clarity and completeness of environment, dependency, and system-level configuration.
    \item \textbf{Versioning Rigor.} Explicit, pinned versions for datasets, models, pipelines, libraries, and tools.
    \item \textbf{Execution Fidelity.} Presence of runnable pipelines, automated scripts, or containers enabling end-to-end replication.
    \item \textbf{Legal Openness.} Licensing clarity for data, models, and code to legally permit reuse and redistribution.
\end{enumerate}

\subsubsection{Four Maturity Levels (RMM-Tiers)}
Based on these axes, we distill four maturity tiers that reflect the empirical distribution of our corpus and provide concrete review criteria.

    \paragraph{RMM-0: Minimal Reproducibility (Low Maturity)} 
    Artifacts are missing, inaccessible, or only partially available. Environment specifications are vague or absent, versions are unpinned, reproduction depends heavily on implicit assumptions. Multiple reproducibility smells can co-occur. 
    
    \subparagraph{Reviewer checklist}
    \begin{itemize}[label=$\square$]
        \item Are any artifacts inaccessible or missing?
        \item Are the environmental details insufficient to recreate the setup?
        \item Are versions unspecified or floating?
        \item Does legal ambiguity prevent reuse?
    \end{itemize}

    \paragraph{RMM-1: Operational Reproducibility (Intermediate Maturity)}
    Artifacts are functional at publication time but vulnerable to dependency drift. Instructions exist, but require manual reconstruction. Containers or scripts may exist, but are incomplete or rely on deprecated APIs. 
    
    \subparagraph{Reviewer checklist}
    \begin{itemize}[label=$\square$]
        \item Are installation steps documented but potentially brittle?
        \item Are versions given but not pinned at all hierarchy levels?
        \item Are pipelines runnable but not fully automated?
        \item Are licenses present but incomplete or unclear?
    \end{itemize}

    \paragraph{RMM-2: Durable Reproducibility (High Maturity)}
    Artifacts are fully runnable, containerized, versioned, and legally open. All five axes are satisfied. Outputs can be reproduced end-to-end with no manual inference. These exemplars (13.3\% of our corpus) demonstrate reproducibility engineered for long-term integrity. 
    
    \subparagraph{Reviewer checklist}
    \begin{itemize}[label=$\square$]
        \item Is a containerized or automated pipeline provided?
        \item Are all dependencies pinned, including OS, libraries, and models?
        \item Are datasets and models persistently hosted with stable identifiers?
        \item Are all components licensed for reuse?
    \end{itemize}

    \paragraph{RMM-3: Independently Verified Reproducibility (Very High Maturity)} 
This tier reflects a level of reproducibility that extends beyond the authors’ own artifacts. An external reviewer or independent research group has successfully re-executed the full experimental pipeline and reproduced the key findings using the provided artifacts without modification. Because this tier requires substantial effort outside the standard publication workflow, we treat it as an \textit{optional}, evaluation-driven extension analogous to the ACM \emph{Results Reproduced} badge rather than a requirement for authors at submission time.

\subparagraph{Reviewer checklist}
\begin{itemize}[label=$\square$]
    \item Can the results be reproduced \emph{exactly} using only the provided artifacts and instructions?
    \item Was the reproduction performed independently, without author intervention?
    \item Are all outputs (tables, figures, metrics) consistent with those reported in the paper?
    \item Does the reproduction process complete successfully in a clean environment?
\end{itemize}



\subsubsection{Axes by Tier Criteria}

Table~\ref{tab:rmm_axes_levels} summarizes how each axis is instantiated at different maturity levels. \textbf{RMM-0} and \textbf{RMM-1} roughly distinguish between missing or brittle artifacts and those that are merely ``momentarily functional'', while \textbf{RMM-2} and \textbf{RMM-3} represent durable, independently verifiable reproducibility.

\begin{table}[!htbp]
\centering
\footnotesize
\caption{Summary of RMM axes and indicative criteria per maturity level. RMM-3 assumes all RMM-2 criteria plus independent re-execution by a third party.}
\label{tab:rmm_axes_levels}
\begin{tabular}{l p{2.9cm} p{2.9cm} p{2.9cm}}
\toprule
\textbf{Axis} & \textbf{RMM-0: Minimal} & \textbf{RMM-1: Operational} & \textbf{RMM-2: Durable} \\
\midrule
Accessibility &
No or fragmented artifacts; links missing, private, or ephemeral. &
Artifacts hosted but with fragile links (personal cloud, ad-hoc URLs); partial coverage of code/data/models. &
Artifacts versioned and persistently hosted (e.g., DOI, archival repository) with clear mapping to experiments. \\ \midrule
Environment Specification &
Environment details largely absent; no explicit hardware/software requirements. &
Basic installation steps documented, but incomplete dependency lists and implicit assumptions about the platform. &
Complete, machine-readable environment specification (e.g., lockfiles, container specs) covering OS, libraries, and hardware. \\\midrule
Versioning Rigor &
Floating versions or unspecified model/dataset snapshots; reproducibility depends on latest defaults. &
Some versions documented (e.g., major library versions), but not consistently pinned across the stack. &
Pinned versions at all hierarchy levels (datasets, models, frameworks, CUDA, etc.), with change logs or manifests. \\\midrule
Execution Fidelity &
No runnable scripts; only high-level prose or pseudo-code. &
Pipelines runnable with manual effort (e.g., several shell commands, manual downloads, ad-hoc scripts). &
End-to-end, automated pipelines or containers that re-create results from scratch with minimal manual steps. \\\midrule
Legal Openness &
Licenses absent, incompatible, or ambiguous; reuse is legally risky. &
Some components licensed, but coverage incomplete or terms unclear for key artifacts (e.g., models, datasets). &
Explicit, compatible licenses for all artifacts required to reproduce results, enabling legal reuse and redistribution. \\
\bottomrule
\end{tabular}
\end{table}

By treating reproducibility as a set of measurable, enforceable criteria, the RMM provides a structured path toward sustainable, verifiable LLM-for-SE research and extends current badging systems with a more holistic maturity spectrum.

\section{Discussion}
\label{sec:discussion}
In this section, we discuss the temporal trends and policy implications of the reproducibility smells. We also discuss how using open and proprietary models is contributing to the issue.

\subsection{Temporal Trends and Policy Implications}

The temporal data (Figure \ref{fig:rq2_smell_trends}) reveal a mixed trajectory. Some reproducibility issues, especially code executability, have improved since 2023, suggesting growing community awareness. However, the occurrence of \category{Access and Legal} and \category{Versioning} smells increased, driven by the usage of proprietary APIs and the absence of fixed model identifiers. In parallel, \category{Environment and Tooling} issues persist at a steady baseline, showing that environment reproducibility remains an issue.

These patterns indicate a structural rather than individual challenge. As LLM-for-SE research increasingly depends on commercial ecosystems, even diligent authors face obstacles in releasing reusable artifacts. This shift from open academic tooling to closed proprietary services calls for policy interventions at the venue level. Artifact-evaluation and reproducibility-badging programs should adapt to this evolving landscape by requiring disclosure of API version, access date, and decoding parameters for any proprietary model, inclusion of at least one open-source baseline to preserve replicability even if commercial endpoints disappear. Moreover, archival of prompt templates, inference settings, and sample interaction logs for LLM-based experiments should also be made a requirements alongside with explicit licensing statements for all code, data, and model artifacts.

\subsection{Open vs.\ Proprietary Model Regimes}
\label{subsec:open_vs_proprietary}

We observed that the reproducibility gap mirrors the divide between \textbf{open} and \textbf{proprietary} model regimes. Open models (\eg LLaMA \cite{grattafiori2024llama3herdmodels}, Mistral \cite{jiang2023mistral7b}) enable complete replication pipelines but require robust environment control and version pinning. Proprietary models (\eg GPT \cite{chatgpt}, Claude \cite{Anthropic2024Claude3}) offer convenience but introduce opacity and temporal drift, as API outputs can change silently. Our work confirms that access/legal and versioning smells are overwhelmingly concentrated in the latter period (post-2023).

To mitigate this, a \textbf{\textit{dual-track reproducibility framework}} can be adopted, in which proprietary-model studies should include at least one open-source baseline. In contrast, open-model studies must ensure full environmental encapsulation and version transparency. Journals and conferences should adapt review rubrics to recognize these distinct reproducibility periods rather than enforcing a uniform standard.

\section{Threats to Validity}
\label{sec:threats}

\paragraph{Internal validity} During the manual data collection and annotation process for reproducibility smells, some information may have been overlooked due to the large scale of the study involving 640 papers and their corresponding repositories. To minimize potential omissions and ensure consistency, two authors with prior experience in artifact evaluation cross-checked the annotations. This expertise and verification process help reduce the likelihood of missing relevant reproducibility information.

\paragraph{External validity} Although the dataset analyzed in this study is large and representative of top-tier venues, it may not fully capture reproducibility practices across the broader research landscape. In particular, our focus on highly visible academic venues such as ICSE, FSE, ASE, and major ML/NLP conferences might bias the results toward research that already follows more formalized artifact and documentation practices. Reproducibility behaviors in non-top-tier venues, workshops, industry-focused conferences, or grey literature (\eg preprints, technical reports) may differ significantly, potentially involving fewer formal artifact evaluation processes, less consistent documentation, or different types of legal and access barriers. However, our findings from top-tier venues provide a strong and conservative baseline. If reproducibility issues are prevalent even in venues that typically enforce stronger standards and review processes, similar or more severe issues are likely to exist in less formal publication settings.

\section{Related Work}
\label{sec:related}
\subsection{Large Language Models for Software Engineering (LLM-for-SE)}

The rapid evolution of large language models (LLMs) has catalyzed a new wave of research in software engineering (SE), giving rise to the emerging field commonly referred to as LLM-for-SE. Unlike earlier program analysis or deep learning for SE approaches, LLM-for-SE leverages general-purpose foundation models such as GPT-4, Codex, and Code Llama to automate or assist in a broad spectrum of SE tasks~\cite{Sallou2024,Baltes2025}. These tasks span the entire software development lifecycle, including code generation, test synthesis, debugging, program comprehension, and software documentation.

Early empirical studies demonstrated that pretrained language models could learn code semantics and structure sufficiently to outperform traditional baselines on code summarization~\cite{Magnusson2023}, code search~\cite{codebert}, and completion tasks~\cite{izadi2022codefill,kim2021code,svyatkovskiy2021fast}. This has encouraged the integration of LLM-powered tools into mainstream development workflows (\eg GitHub Copilot). However, these models differ significantly from earlier task-specific ML approaches: they are often hosted behind APIs, trained on massive and largely opaque datasets, and fine-tuned continuously, which introduces significant reproducibility challenges~\cite{Sallou2024}. Our work focuses on the current situation of the reproducibility of software engineering tasks using large language models.

\subsection{Reproducibility in Software Engineering and Machine Learning}

Reproducibility has emerged as a critical concern across science and engineering in recent years. Surveys indicate that over 70\% of researchers have failed to reproduce another group's findings, highlighting a cross-disciplinary “reproducibility crisis”~\cite{Baker2016}. In software engineering (SE), reproducibility is foundational for building trust in empirical results and enabling independent verification~\cite{Krishnamurthi2015}. Krishnamurthi and Vitek famously argued that repeatability is the "real software crisis," urging the community to treat reproducibility as a core value in SE research. Early studies of SE experiments uncovered substantial barriers to repetition. For example, González-Barahona and Robles identified difficulties in reusing data retrieval tools, prompting calls for better reproducibility practices~\cite{GonzalezBarahona2012}, and their follow-up work a decade later confirmed persistent gaps~\cite{Robles2023}.

Top SE venues have responded by institutionalizing artifact sharing and evaluation processes to improve reproducibility. Conferences such as ICSE, FSE, ASE, and ISSTA now regularly include Artifact Evaluation Committees and award reproducibility badges for papers that provide accessible code and data~\cite{Hermann2020,Timperley2021,Li2024}. This cultural shift has led to a steady rise in artifact availability over the past decade. Qualitative studies have explored the impact of these initiatives. Hermann \etal~examined community expectations for artifacts~\cite{Hermann2020}, while Timperley \etal~analyzed artifact evolution trends~\cite{Timperley2021}. Still, domain-specific hurdles remain; for example, replication in NLP-for-Requirements-Engineering required significant re-implementation effort due to under-specified models and data preparation steps~\cite{Abualhaija2024}.

The machine learning (ML) and natural language processing (NLP) communities face similar concerns. Reproducibility issues in ML are often linked to missing code, opaque experimental setups, and insufficient reporting~\cite{pineau2021improving,Magnusson2023}. Even when code is shared, it may lack critical preprocessing instructions or fixed hyperparameters, leading to inconsistent results across environments. Liu \textit{et al.} demonstrated that many SE deep learning results were inflated or unstable across runs, highlighting the need for multiple runs and better reporting~\cite{Liu2022}. To address such problems, venues like NeurIPS and ICLR introduced reproducibility checklists, increasing the proportion of papers sharing code and methodological details~\cite{pineau2021improving}, while ACL introduced its Responsible NLP Checklist to improve reporting standards~\cite{Magnusson2023}. 


Hassan \etal~\cite{HASSAN2025112327} conducted a systematic literature review on reproducibility debt in scientific software, introducing a seven-dimensional taxonomy encompassing data, code, documentation, processes, human factors, legal constraints, and versioning. Their work provides a high-level conceptual framework to understand reproducibility breakdowns in computational sciences. In contrast, our study targets a rapidly changing and highly specialized domain, LLM-for-SE and develops a reproducibility smell taxonomy grounded in empirical evidence drawn from 640 papers published between 2020 and 2025. While inspired by Hassan \etal's structure, our taxonomy is to capture domain-specific reproducibility pitfalls in LLM workflows, evaluation practices, and software engineering tooling. Furthermore, our analysis is not limited to conceptual mapping but includes a systematic quantification of reproducibility smells, offering actionable insights for both researchers and publication venues in the LLM-for-SE space.

\subsection{Reproducibility Challenges in LLM-for-SE Research}
LLM-for-SE research inherits reproducibility challenges from both domains. A key issue is the reliance on closed-source or continuously evolving models (e.g., GPT-4 or Codex), where researchers cannot control or replicate the exact model state~\cite{Sallou2024}. Experiments relying on such APIs are vulnerable to model updates and access restrictions, limiting repeatability. Further, complex dependencies and environmental configurations introduce additional failure points, especially when model or dataset versioning is incomplete. Prompt design and generation parameters significantly impact results, with small variations yielding divergent outcomes~\cite{siddiq2024benchmarks}.

Addressing these challenges requires both cultural and technical interventions. In SE, artifact evaluation and open science initiatives have been widely adopted~\cite{Li2024}, while in ML/NLP, reproducibility checklists and data-sharing incentives have shown measurable impact~\cite{pineau2021improving,Magnusson2023}. Beyond policies, tooling such as containerization (\eg Docker) and data/model versioning (\eg DVC) has become central to reproducible experimentation. Emerging guidelines for LLM-based SE studies advocate documenting prompts, model versions, and intermediate artifacts~\cite{Baltes2025,Wolter2025} to ensure verifiable and transparent research practices. Building on these insights, our work provides a large-scale empirical assessment of reproducibility in LLM-for-SE papers and proposes actionable guidelines informed by both SE artifact evaluation culture and ML/NLP reproducibility initiatives.

Williams \etal~\cite{williams2025reflectingempiricalsustainabilityaspects} provide a high-level assessment of LLM-based SE research at ICSE, focusing on model and benchmark usage, contamination risks, and broad replicability indicators such as whether artifacts or badges are present. In contrast, our study conducts a cross-venue, multi-year analysis and introduces a fine-grained taxonomy of reproducibility smells, quantifying their prevalence, co-occurrence, and evolution.

\subsection{Critical Reviews of Software Engineering Research Practices}
A growing body of meta-research within empirical software engineering has highlighted foundational methodological weaknesses and reporting deficiencies. Stol \etal~\cite{Stol2016} examine 98 SE articles that mention or claim to use the Grounded Theory (GT) method. They find that only 16 of the 98 provide detailed accounts of their research procedures, and many studies show “method-slurring” (mixing GT techniques without coherent adherence). The paper offers guidelines for the conduct and reporting of GT in SE. Baltes \etal~\cite{Baltes2022} perform a critical review of sampling in recent high-quality SE empirical studies and report three major findings: (1) random sampling is rare; (2) sophisticated sampling strategies are extremely rare; (3) sampling, representativeness, and randomness are often misunderstood. They argue that SE research suffers from a generalisability crisis. Huang \etal~\cite{Huang2018} review how qualitative research synthesis (QRS) is carried out in SE, identifying several methodological deficiencies (\eg inconsistent definitions, weak reporting of how primary studies were selected and synthesised). While these works focus on critical reviews of various practices in software engineering research, our work focuses on the reproducibility of LLM-for-SE research.

\section{Conclusion}
\label{sec:conclusion}
Reproducibility remains a critical and under-addressed challenge in large language model (LLM)-based software engineering (SE) research. Through a systematic analysis of \textbf{640 papers} across premier SE, ML, and NLP venues, we observed persistent reproducibility smells, particularly in code execution pipelines, versioning rigor, and access and legal constraints. While traditional SE research has benefited from structured artifact evaluation and open-science initiatives, LLM-for-SE introduces domain-specific barriers, including dependence on proprietary APIs, opaque prompt designs, non-deterministic inference behaviors, and rapidly evolving model and package ecosystems.

Our temporal results reveal modest progress in recent years, driven in part by reproducibility policies at top-tier venues. However, even as artifact evaluation badges become more common, they do not consistently guarantee execution fidelity or long-term replicability. These findings highlight a gap between \emph{artifact availability} and \emph{artifact maturity}. To bridge this gap, we introduce a \textbf{Reproducibility Maturity Model (RMM)} that moves beyond binary badge systems toward multi-dimensional assessments encompassing accessibility, environment specification, version control, execution fidelity, and legal openness.

We release our dataset, taxonomy, and analysis framework to support future benchmarking and automated smell detection efforts. By embracing maturity-oriented reproducibility practices, the LLM-for-SE community can ensure that research remains verifiable, buildable, cumulative, and trustworthy.

\bibliography{references,papers}

\end{document}